\documentclass[%
superscriptaddress,
nofootinbib,
 amsmath,amssymb,
 aps,
twocolumn
]{revtex4-2}

\usepackage{graphicx}
\usepackage{dcolumn}
\usepackage{bm}
\usepackage{hyperref}
\usepackage{multirow}

\usepackage{bbold}

\newcommand{\beq}{\begin{equation}}
\newcommand{\eeq}{\end{equation}}
\newcommand{\bea}{\begin{eqnarray}}
\newcommand{\eea}{\end{eqnarray}}
\newcommand{\nn}{\nonumber}

\newcommand{\Tr}{\mathop{\rm Tr}}

\newcommand{\third}{\frac 13}

\newcommand{\sixth}{\frac 16}

\newcommand{\Slash}[1]{{\ooalign{\hfil#1\hfil\crcr\raise.167ex\hbox{/}}}}

\begin{document}


\title{Good flavor search in $SU(5)$: a machine learning approach}

\author{Fayez Abu-Ajamieh}
    \email{fayez@soknengineering.com}
    \affiliation{Formerly
 Center for High Energy Physics,
 Indian Institute of Science,
 Bangalore 560012, Karnataka, India
}
\author{Shinsuke Kawai}
	\email{kawai@sci.kj.yamagata-u.ac.jp}
	\affiliation{%
    Faculty of Science, Yamagata University,
    1-4-12 Kojirakawa-machi, Yamagata, 990-8560 Japan
}%
\author{Nobuchika Okada}
 \email{okadan@ua.edu}
    \affiliation{%
    Department of Physics and Astronomy, University of Alabama, Tuscaloosa, Alabama, AL35487 USA
}%


\date{\today}

\begin{abstract}

We revisit the fermion mass problem of the $SU(5)$ grand unified theory using machine learning techniques.  
The original $SU(5)$ model proposed by Georgi and Glashow is incompatible with the observed fermion mass spectrum. 
Two remedies are known to resolve this discrepancy, one is through introducing a new interaction via a 45-dimensional field, and the other via a 24-dimensional field.  
We investigate which modification is more beautiful, defining the beauty as proximity to the original Georgi-Glashow $SU(5)$ model. 
Our analysis shows that, in both supersymmetric and non-supersymmetric scenarios, the model incorporating the interaction with the 24-dimensional field is more beautiful under this criterion.  
We then generalise these models by introducing a continuous parameter $y$, which takes the value 3 for the 45-dimensional field and 1.5 for the 24-dimensional field.
Numerical optimisation reveals that $y \approx 0.8$ yields the closest match to the original $SU(5)$ model, indicating that this value corresponds to the most beautiful model according to our definition.

\end{abstract}

\maketitle


\section{\label{sec:intro}Introduction}

Machine learning has become a key technology in modern scientific research, demonstrating utility also in theoretical investigations.  
Its application is particularly powerful when a theoretical model with a vast parameter space is confronted with extensive experimental data \cite{ATLAS:2012yve,CMS:2012qbp,Abbott:2016blz}, or when the task involves classifying a large ensemble of theoretical models \cite{He:2017set,Halverson:2019tkf}.  
Even in cases involving a moderately large number of parameters, such as in the Standard Model of particle physics, brute-force numerical analysis is often impeded by the curse of dimensionality, i.e., limitations in practical computational resources. 
In such cases, machine learning techniques have proven to be highly effective \cite{Matchev:2024ash} (see also \cite{Harvey:2021oue,deSouza:2022uhk,Nishimura:2020nre,Romao:2024gjx,Nishimura:2024apb}).

We revisit the fermion mass problem in the context of grand unified theories (GUTs).  
The original $SU(5)$ GUT proposed by Georgi and Glashow \cite{Georgi:1974sy}, a prototypical framework for physics {\em beyond} the Standard Model, organizes the six quarks $u$, $d$, $c$, $s$, $t$, $b$ and the six leptons $e$, $\mu$, $\tau$, $\nu_e$, $\nu_\mu$, $\nu_\tau$ into three families, each comprising the $\bm{10}$ and $\overline{\bm{5}}$ representations of the $SU(5)$ gauge group.  
This structure leads to the following mass relations at the GUT scale:
\begin{align}\label{eqn:gutmassrel}
	m_e = m_d, \quad m_\mu = m_s, \quad m_\tau = m_b.
\end{align}
Fermion mass predictions are then obtained via renormalisation group analysis, using these relations as boundary conditions. 
However, the resulting masses are incompatible with experimental observations; this discrepancy constitutes the fermion mass problem.  
One proposed solution involves a higher-dimensional operator that couples the fermions to the $\bm{24}$-dimensional (adjoint) Higgs field \cite{Ellis:1979fg}, referred to here as the 24-Higgs model.  
Another well-known approach introduces a Higgs field in the $\bm{45}$-dimensional representation of $SU(5)$, which also couples to the fermions \cite{Georgi:1979df}; we refer to this as the 45-Higgs model.  
In both models, new (effective) Yukawa couplings are introduced, and by tuning the additional parametric degrees of freedom, the models can accommodate the observed fermion mass spectrum and flavour mixings.

While both approaches are theoretically well motivated, neither leads to strong predictability.
The (effective) Yukawa coupling, expressed as a $3\times 3$ complex matrix, introduces 18 additional real parametric degrees of freedom, rendering the space of parameters consistent with the observed quark and lepton mass spectrum enormously large.
Within such an expansive parameter space, which set of values is most likely realised in nature?
Prioritising among parameter values that all satisfy observational constraints inevitably involves an element of human subjectivity.
Following \cite{Matchev:2024ash,Kawai:2015ryj}, we refer to the quality that guides such preference as {\em beauty}.
A commonly used principle in practical model selection is {\em Occam’s razor}, which asserts that simplicity is the standard of beauty.
In physics, this is often characterised by larger symmetry and fewer degrees of freedom.
In the context of $SU(5)$ GUTs, the original Georgi–Glashow model may be regarded as the simplest, and hence the most beautiful, yet it is incompatible with observation.
We may therefore consider, within the parameter space of extended GUT models, the proximity to the Georgi–Glashow model as a criterion for beauty.
This choice is particularly well grounded in the case of the 24‑Higgs model, as it involves a Planck‑suppressed operator expected to be small, thereby favouring proximity to the Georgi–Glashow limit.
In \cite{Kawai:2024pws}, we employed machine‑learning techniques for this purpose.
By defining a loss function that quantifies our criterion of beauty, namely, closeness to the original Georgi–Glashow $SU(5)$ model, we demonstrated that gradient-based numerical optimisation, a core tool of machine learning, can be effectively applied in situations where brute‑force parameter scans are impractical.
Using this criterion, we found that the 24‑Higgs model is statistically more beautiful than the 45‑Higgs model.

The aim of the present paper is to extend the study of \cite{Kawai:2024pws} in several directions. 
We first consider grand unification without supersymmetry. 
In \cite{Kawai:2024pws} only the supersymmetric $SU(5)$ scenario was analysed, due mainly to its simplicity. 
Without convincing experimental data supporting supersymmetry thus far, it is reasonable to suppose that nature may well accomplish grand unification without supersymmetry. 
Our analysis shows that the result is, similar to the supersymmetric case, the 24-Higgs model is closer to the original model of Georgi and Glashow, and hence, is more beautiful than the 45-Higgs model. 
Another direction of extension is generalisation beyond the 24-Higgs or the 45-Higgs model.  
We develop a model parametrised by a continuous variable $y$ that takes the value 3 for the 45-Higgs model and 1.5 for the 24-Higgs model. 
We find that the 24-Higgs model is not the most beautiful in this context; it turns out that the most beautiful case corresponds to $y\approx 0.8$.

In next section we review the $SU(5)$ GUT and lay out our notation of the flavour sector of the model. 
Sec. \ref{sec:nonSUSY} presents results of the nonsupersymmetric $SU(5)$ GUT. 
Results of the supersymmetric case are given in Sec. \ref{sec:SUSY}, which partially overlap with the results of \cite{Kawai:2024pws} but includes some technical improvements.
The one-parameter generalised model is introduced and the results are presented in Sec. \ref{sec:ymodel}. 
In Sec. \ref{sec:final} we conclude with comments.

\section{\label{sec:model}Fermion masses in $SU(5)$ GUT}

\subsection{\label{sec:minSU5}GUT mass relation of the minimal $SU(5)$ model}

The $SU(5)$ Georgi-Glashow model \cite{Georgi:1974sy} organises the quarks and leptons of the Standard Model into five and ten dimensional $SU(5)$ fermions $\Psi^i_{\bm{\overline 5}}$ and $\Psi^i_{\bm{10}}$, where $i \in \{1, 2, 3\}$ is the index for the family.  
For the first generation $i=1$, the Standard Model fermions are embedded as
\begin{align}\label{eqn:F5}
	[\Psi^1_{\bm{\overline 5}}]^m=\begin{bmatrix}
		d_1^c,\; d_2^c,\; d_3^c,\; e,\; -\nu_e
	\end{bmatrix},
\end{align}
\begin{align}\label{eqn:F10}
	[\Psi^1_{\bm{10}}]_{mn}=\frac{1}{\sqrt 2}\begin{bmatrix}
		0 & u_3^c & -u_2^c & -u_1 & -d_1\\
		-u_3^c & 0 & u_1^c & -u_2 & -d_2\\
		u_2^c & -u_1^c & 0 & -u_3 & -d_3\\
		u_1 & u_2 & u_3 & 0 & -e^c \\
		d_1 & d_2 & d_3 & e^c & 0
	\end{bmatrix},
\end{align}
and similarly for $i = 2,3$.
The subscripts 1, 2, 3 for the $u$ and $d$ indicate the three colours of the quarks and $m,n,...\in\{1, 2, 3, 4, 5\}$ are the $SU(5)$ indices.
We shall suppress the indices unless they are needed.

Besides the 3 families of fermionic fields $\Psi^i_{\bm{\overline 5}}$ and $\Psi^i_{\bm{10}}$, the $SU(5)$ model consists of a $\bm{24}$ representation gauge field, a $\bm{24}$ representation Higgs field $H_{\bm{24}}$ and a $\bm{5}$ representation Higgs field $H_{\bm{5}}$.
The $H_{\bm{24}}$ field is responsible for GUT symmetry breaking $SU(5)\to SU(3)_c\times SU(2)_L\times U(1)_Y$, whereas the $H_{\bm{5}}$ field is responsible for the electroweak symmetry breaking $SU(2)_L\times U(1)_Y\to U(1)_{EM}$.
The Yukawa part of the Lagrangian reads
\begin{align}\label{eqn:YukawaL}
	y^{5d}_{ij}[\overline{\Psi^i_{\bm{\overline 5}}}]^m[\Psi^j_{\bm{10}}]_{mn}[H_{\bm{\overline 5}}]^n
	&+y^{5u}_{ij}\epsilon^{mnpqr}[\overline{\Psi^i_{\bm{10}}}]_{mn}[\Psi^j_{\bm{10}}]_{pq}[H_{\bm{5}}]_r\nonumber\\
	&\qquad\qquad\qquad\qquad\quad+{\rm h.c.},
\end{align}
where $y^{5u}_{ij}$, $y^{5d}_{ij}$ are the Yukawa coupling matrices and $\epsilon$ is the antisymmetric tensor.
Here in the nonsupersymmetric setup $H_{\bm{\overline 5}}$ is the complex conjugate of $H_{\bm{5}}$.
Upon electroweak symmetry breaking the $H_{\bm{5}}$ Higgs field is assumed to acquire vacuum expectation value
\begin{align}\label{eqn:H5vev}
	\langle [H_{\bm{5}}]_m\rangle=[0, 0, 0, 0, \frac{v_h}{\sqrt 2}]
\end{align}
and the Lagrangian \eqref{eqn:YukawaL} becomes
\begin{align}
	y^{5d}_{ij}\frac{v_h}{\sqrt 2}[\overline{\Psi^i_{\bm{\overline 5}}}]^m[\Psi^j_{\bm{10}}]_{m5}
	&+y^{5u}_{ij}\frac{v_h}{\sqrt 2}\epsilon^{mnpq5}[\overline{\Psi^i_{\bm{10}}}]_{mn}[\Psi^j_{\bm{10}}]_{pq}\nonumber\\
	&\qquad\qquad\qquad\qquad\qquad+{\rm h.c.},
\end{align}
from which the fermion mass matrices are found to be
\begin{align}\label{eqn:M5rel}
	M_u =4\frac{v_h}{\sqrt 2} y^{5u}_{ij},\;\;
	M_d = \frac{v_h}{2} y^{5d}_{ij},\;\;
	M_e = \frac{v_h}{2} y^{5d}_{ji} = M_d^T.
\end{align}
The relations \eqref{eqn:gutmassrel} follow from the last equation of \eqref{eqn:M5rel}.
Using renormalisation group analysis, this prediction at the GUT scale may be compared with the low energy fermion masses that we know by experiments.
It turns out that the third generation relation ($m_\tau=m_b$) is approximately consistent, but the relations for the first and second generations wildly violate the experimental constraints. 
This is the fermion mass problem, indicating that the model needs to be modified.

\subsection{\label{sec:H45}The 45-Higgs model}

A well known solution \cite{Georgi:1979df} to the fermion mass problem is to include $\bm{\overline{45}}$ representation Higgs field\footnote{
In the supersymmetric case (Sec.~\ref{sec:SUSY}), this needs to be accompanied by the partner $H_{\bm{45}}$ in a vectorlike pair $(\bm{45},\bm{\overline{45}})$, which satisfies 
$[H_{\bm{45}}]^{m}_{np}=-[H_{\bm{45}}]^{m}_{pn}$, $[H_{\bm{45}}]^{m}_{mp}=0$.
A gauge invariant term
$\epsilon^{mnprs}[F^i_{\bm{10}}]_{mn}[F^j_{\bm{10}}]_{pq}[H_{\bm{45}}]^{q}_{rs}$
can arise, but it is not important for our discussions below.
}
$H_{\bm{\overline{45}}}$ satisfying
$[H_{\overline{\bm{45}}}]^{np}_m=-[H_{\overline{\bm{45}}}]^{pn}_m$, 
$[H_{\overline{\bm{45}}}]^{mp}_m=0$, 
that has $SU(3)_c\times U(1)_Y$-invariant vacuum expectation value
\begin{align}\label{eqn:H45vev}
	\langle[H_{\overline{\bm{45}}}]^{n5}_m\rangle = v_{45}\;{\rm diag}(1, 1, 1, -3, 0).
\end{align}
An $SU(5)$ gauge invariant Yukawa term
\begin{align}\label{eqn:H45lag}
	y^{45d}_{ij}[\Psi^i_{\bm{\overline 5}}]^m[\Psi^j_{\bm{10}}]_{np}[H_{\overline{\bm{45}}}]^{np}_m,
\end{align}
with $3\times 3$ complex Yukawa matrix $y^{45d}_{ij}$, then gives additional terms to the mass matrices.
Including such contributions the mass matrices now read
\begin{align}\label{eqn:M45rel}
	M_u =& 2\sqrt 2 v_h y^{5u}_{ij},\nonumber\\
	M_d =& \frac{v_h}{2} y^{5d}_{ij}+ \frac{v_{45}}{\sqrt 2}\, y^{45d}_{ij},\nonumber\\
	M_e =& \frac{v_h}{2} y^{5d}_{ji}-3\frac{v_{45}}{\sqrt 2}\, y^{45d}_{ji}.
\end{align}
The fermion mass relations are seen to be modified by the new terms, and the observed fermion mass spectrum at low energy can be recovered by adjusting $y^{5d}_{ij}$ and $y^{45d}_{ij}$.
For convenience let us write
\begin{align}
	M_5\equiv  \frac{v_h}{2} y^{5d}_{ij},\quad
	M_{45}\equiv \frac{v_{45}}{\sqrt 2}\, y^{45d}_{ij}.
\end{align}
Then the second and third lines of \eqref{eqn:M45rel} are concisely
\begin{align}\label{eqn:MM45}
	M_5 &= \frac 14 (3M_d+M_e^T),\nonumber\\
	M_{45} &= \frac 14 (M_d-M_e^T).
\end{align}
 

\subsection{\label{sec:H24}The 24-Higgs model}%

An alternative solution to the fermion mass problem is to stay within the same field contents of the Georgi-Glashow model and take into account the effects of higher dimensional operator \cite{Ellis:1979fg}
\begin{align}\label{eqn:H24lag}
	\frac{y^{24d}_{ij}}{m_{\rm P}}[\overline{\Psi^i_{\bm{\overline 5}}}]^m [H_{\bm{24}}]_m{}^n [\Psi^j_{\bm{10}}]_{np} [H_{\bm{\overline 5}}]^p+{\rm h.c.}, 
\end{align}
which is a $SU(5)$ gauge singlet. 
Here, $y^{24d}_{ij}$ is a new Yukawa-like coupling and $m_{\rm P}$ is the Planck mass.
The $H_{\bm{24}}$ field acquires expectation value
\begin{align}\label{eqn:H24vev}
	H_{\bm{24}}=v_{24}\,{\rm diag}(2,2,2,-3,-3)
\end{align}
upon GUT symmetry breaking.
The $H_{\bm{5}}$ field acquires expectation value \eqref{eqn:H5vev} below the electroweak scale, and here $H_{\bm{\overline 5}}$ is the complex conjugate of $H_{\bm{5}}$.
The fermion mass matrices including the effects from the higher dimensional term \eqref{eqn:H24lag} then becomes,
\begin{align}\label{eqn:M24rel}
	M_u =& 2\sqrt 2 v_h y^{5u}_{ij},\nonumber\\
	M_d =& \frac{v_h}{2} y^{5d}_{ij}+ \frac{v_{24}}{m_{\rm P}}v_h\, y^{24d}_{ij},\nonumber\\
	M_e =& \frac{v_h}{2} y^{5d}_{ji}- \frac{3}{2}\frac{v_{24}}{m_{\rm P}}v_h\, y^{24d}_{ji}.
\end{align}
Tuning $y^{5d}_{ij}$ and $y^{24d}_{ij}$ the fermion mass spectrum observed at low energy is recovered via renormalisation group flow.
Defining 
\begin{align}
M_{24} \equiv \frac{v_{24}}{2m_{\rm P}}v_h\, y^{24d}_{ij},
\end{align}
the last two lines of \eqref{eqn:M24rel} are written
\begin{align}\label{eqn:MM24}
	M_5 &= \frac 15 (3M_d+2M_e^T),\nonumber\\
	M_{24} &= \frac 15 (M_d-M_e^T). 
\end{align}

\subsection{\label{sec:Loss}Beauty of models and loss function}%

We often encounter situations in physics where a configuration predicted by a simple mathematical principle (such as maximal symmetry) is not realised in nature, but instead experimental data exhibit slight deviations from it.
The $SU(5)$ GUT under consideration clearly exemplifies this feature.
In the parlance of \cite{Matchev:2024ash,Kawai:2024pws}, the model of maximal {\em beauty}, represented here by the Georgi-Glashow model, is at odds with the {\em truth}, namely the fermion mass spectrum observed in experiments.
Our objective is to identify parameter values that optimally reconcile beauty with truth, that is, to locate a {\em good} point in the parameter space as close as possible to the Georgi-Glashow model while remaining consistent with experimental constraints.

This conceptually straightforward task presents two technical challenges. 
The first concerns the size of the parameter space:
due to its relatively high dimensionality, a brute-force scan is computationally infeasible. 
The central result of this paper, as we will demonstrate, is that this difficulty can be circumvented by employing machine learning techniques.
More precisely, for the analysis of the GUT models we use parameter sampling combined with gradient-based numerical optimisation, which is one of the key tools in modern machine learning.
The second challenge lies in defining the criterion of beauty, that is, quantifying the proximity to the Georgi-Glashow model within the parameter space. 
To address this, we examine the structure of the mass matrices that underlie the fermion mass problem.
The Georgi-Glashow model predicts $M_d - M_e^T=0$, corresponding to the third equation in \eqref{eqn:M5rel}, which is experimentally excluded. 
Indeed, as seen from \eqref{eqn:MM45} and \eqref{eqn:MM24}, the modifications introduced via the 45-Higgs and 24-Higgs models are specifically designed to shift the value of $M_d - M_e^T$ away from zero.
Our goal is to identify parameter values that are consistent with experimental data while remaining as close as possible to the point $M_d - M_e^T=0$.
To this end, we propose using $\det(M_d - M_e^T)$, appropriately normalised, as a measure of distance from the point of maximal beauty. 
Note that the determinant is a convenient measure as it is invariant under unitary transformations.
Specifically, as a dimensionless measure of this distance, we define
\begin{align}\label{eqn:Ldef}
	L = \left|\frac{\det(M_d - M_e^T)}{\det M_5}\right|,
\end{align}
and aim to minimise it subject to experimental constraints.
The quantity \eqref{eqn:Ldef} is regarded as a {\em loss function}. 
We shall use standard techniques of machine learning to optimise it in Secs.\ref{sec:nonSUSY}, \ref{sec:SUSY}.

\subsection{\label{sec:Params}Parametrisation of the mass matrices}%

As experimental inputs, we demand the fermion masses (in GeV) and the Cabibbo–Kobayashi–Maskawa (CKM) parameters take following values at energy scale $M_Z=91.2$ GeV \cite{Ohlsson:2018qpt}:
\begin{widetext}
\begin{align}\label{eqn:fmasses}
  	m_u(M_Z) &= 0.00127,  & m_c(M_Z)   &= 0.634,  & m_t(M_Z)    &= 171, \nonumber\\
	m_d(M_Z) &= 0.00271,  & m_s(M_Z)   &= 0.0553, & m_b(M_Z)    &= 2.86, \nonumber\\
	m_e(M_Z) &= 0.000487, & m_\mu(M_Z) &= 0.103,  & m_\tau(M_Z) &= 1.75, 
\end{align}
\begin{align}\label{eqn:CKMvalues}
	s_{12}(M_Z)&=0.225,   & s_{23}(M_Z)&=0.0411, &
	s_{13}(M_Z)&=0.00357, & \delta_{CP}(M_Z) &= 1.24.
\end{align}
\end{widetext}
The CKM parameters are in the standard parametrisation (i.e. the CKM matrix is the form of \eqref{eqn:CKMlike} below).
These are the centre values; 
renormalisation group analysis including uncertainties can be found e.g. in \cite{Antusch:2025fpm}, but changing the input values within 1-$\sigma$ does not alter our conclusions below.
The strong CP phase is experimentally zero-consistent.

We will parametrise observationally unconstrained degrees of freedom in the Yukawa sector while ensuring experimental consistency. 
Recall first that an arbitrary $3\times 3$ complex matrix can be diagonalised by two unitary matrices $V$ and $U$.
Then the up-type quark, down-type quark, and charged lepton Yukawa matrices are diagonalised as
\begin{align}
	\frac{v_h}{\sqrt 2}V_u^\dag y^u U_u = {\rm diag}(m_u, m_c, m_t)\equiv D_u,\nonumber\\
	\frac{v_h}{\sqrt 2}V_d^\dag y^d U_d = {\rm diag}(m_d, m_s, m_b)\equiv D_d,\nonumber\\
	\frac{v_h}{\sqrt 2}V_e^\dag y^e U_e = {\rm diag}(m_e, m_\mu, m_\tau)\equiv D_e.
\end{align}
We will not discuss the neutrino Dirac Yukawa coupling, assuming that the seesaw scale is not far from the GUT scale and hence the neutrino sector contributions are negligible.
The CKM matrix is
\begin{align}\label{eqn:CKMdef}
	V_{\rm CKM}=V_u^\dag V_d,
\end{align}
and the mass matrices $M_u$, $M_d$, $M_e$, appearing in \eqref{eqn:M5rel}, \eqref{eqn:M45rel} and \eqref{eqn:M24rel}, are 
\begin{align}\label{eqn:Mdiag}
	M_u\equiv &\frac{v_h}{\sqrt 2}y^u = V_uD_uU_u^\dag,\nonumber\\
	M_d\equiv &\frac{v_h}{\sqrt 2}y^d = V_dD_dU_d^\dag,\nonumber\\
	M_e\equiv &\frac{v_h}{\sqrt 2}y^e = V_eD_eU_e^\dag.
\end{align}
We choose the unitary matrices $V_u$, $U_u$, $V_e$ and $U_e$ so that $M_u$ and $M_e$ are diagonal:
\begin{align}\label{eqn:MuMe}
	M_u &= D_u = {\rm diag}(m_u, m_c, m_t),\nonumber\\
	M_e &= D_e = {\rm diag}(m_e, m_\mu, m_\tau),
\end{align}
which is always possible. 
Then $M_d$ of \eqref{eqn:Mdiag} is not diagonalisable.
It is expressed, using $V_d=V_uV_{\rm CKM}$ from \eqref{eqn:CKMdef}, as
\begin{align}\label{eqn:Md}
	M_d = V_uV_{\rm CKM}D_dU_d^\dag,
\end{align}
where $V_u$ is diagonal and $U_d^\dag$ is unitary.
An arbitrary $3\times 3$ unitary matrix has 9 real degrees of freedom and it can be parametrised as
\begin{align}\label{eqn:UnitaryParam}
	&U(\phi_0,\phi_1,\phi_2,\theta_1,\theta_2,\delta, \theta_3, \chi_1, \chi_2)\nonumber\\
	&=e^{i\phi_0} e^{i(\phi_1\lambda_3+\phi_2\lambda_8)}
	R(\theta_1,\theta_2,\delta,\theta_3) e^{i(\chi_1\lambda_3+\chi_2\lambda_8)},
\end{align}
where $\lambda_3={\rm diag}(1, -1, 0)$, $\lambda_8={\rm diag}(1, 1, -2)/\sqrt 3$ and
\begin{align}\label{eqn:CKMlike}
	&R(\theta_1,\theta_2,\delta,\theta_3)\\
	&=\begin{pmatrix}1&0&0\\ 0&c_{1}&s_{1}\\ 0&-s_{1}&c_{1}\end{pmatrix}
	\begin{pmatrix}c_{2}&0&s_{2}e^{-i\delta}\\ 0&1&0\\ -s_{2}e^{i\delta}&0&c_{2}\end{pmatrix}
	\begin{pmatrix}c_{3}&s_{3}&0\\ -s_{3}&c_{3}&0\\ 0&0&1\end{pmatrix}\nonumber
\end{align}
is a CKM-like matrix, with $s_i\equiv\sin\theta_i$ and $c_i\equiv\cos\theta_i$.

We are concerned with the physics at the unification scale $M_U$ and will parametrise the mass matrices as follows.
The mass matrices $M_u$ and $M_e$ are taken diagonal,

\begin{align}
	M_u(M_U) = (m_u(M_U), m_c(M_U), m_t(M_U)),\crcr
	M_e(M_U) = (m_e(M_U), m_\mu(M_U), m_\tau(M_U)),
\end{align}
namely, the fermion masses at the unification scale.
In general, they differ from the low energy values \eqref{eqn:fmasses} due to renormalisation.
The mass matrix $M_d(M_U)$ is not diagonal and is expressed as \eqref{eqn:Md}.
Parametrising the unitary matrix $U_d^\dag$ as \eqref{eqn:UnitaryParam} and rearranging the diagonal elements, it is expressed using 11 real parameters $x_0, \cdots, x_{10}$, $0\leq x_i<2\pi$, as
\begin{align}\label{eqn:MdParam}
	&M_d(M_U)\nonumber\\
	&= e^{ix_0}\,e^{i(x_1\lambda_3+x_2\lambda_8)}\,V_{\rm CKM}(M_U)\,D_d(M_U)\nonumber\\
	&\;\times e^{i(x_3\lambda_3+x_4\lambda_8)} R(x_5, x_6, x_7, x_8)\, e^{i(x_9\lambda_3+x_{10}\lambda_8)},
\end{align}
where
\begin{align}
	D_d(M_U) = {\rm diag}(m_d(M_U), m_s(M_U), m_b(M_U)).
\end{align}
These fermion masses and the CKM matrix $V_{\rm CKM}(M_U)$ must be evaluated at the GUT scale $M_U$.
They are obtained by solving renormalisation group equations, employing the low-energy values \eqref{eqn:fmasses}, \eqref{eqn:CKMvalues} as boundary conditions.
The renormalisation group flow is dependent on the scenario of GUT (e.g. supersymmetric or non-supersymmetric), as we elaborate on the details in the subsequent sections.

The bounds on the neutron electric dipole moment \cite{Abel:2020pzs} indicates smallness of the strong CP parameter $|\overline\theta|\lesssim 10^{-10}$, which is related to the (topological) QCD angle $\theta_{\rm QCD}$ and the fermion mass matrices as
\begin{align}\label{eqn:thetabar}
	\overline\theta=\theta_{\rm QCD}-\arg\det(M_uM_d).
\end{align}
We assume the two terms of \eqref{eqn:thetabar} separately vanish. 
As we chose $M_u$ to be real and diagonal in \eqref{eqn:MuMe}, $\arg\det M_u$ is identically zero. 
Inspecting each factor of $M_d(M_U)$ in \eqref{eqn:MdParam}, it turns out that a nontrivial phase can arise only from the first factor.
Then the condition for the vanishing CP angle
\begin{align}
	\arg\det(e^{ix_0}{\rm I}_{3\times 3}) = \arg(e^{3ix_0}) = 0
\end{align}
restricts $x_0$ to three possible values,
\begin{align}
	x_0 = 0, \quad\frac 23 \pi, \quad\frac 43 \pi.
\end{align}
Thus, considering the low energy data, we investigate the flavour sector of the $SU(5)$ model parametrised by 10 continuous variables $0\leq x_i<2\pi$, $i=1,\cdots,10$ and three possible cases $x_0 = 0, \frac 23 \pi, \frac 43 \pi$.

\begin{figure}[t]
\includegraphics[width=95mm]{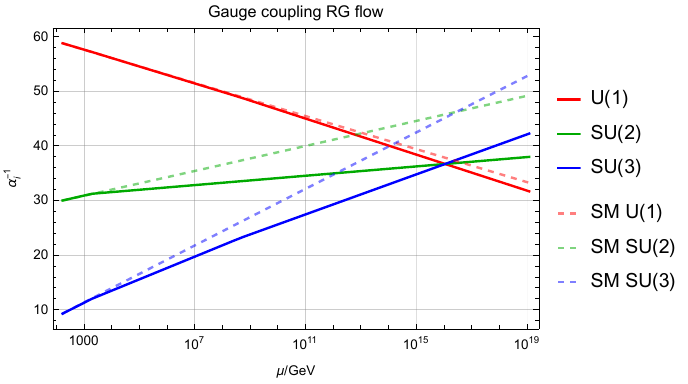}
\caption{\label{fig:NonSUSY_gRG} 
Renormalisation group flow of $\alpha_i^{-1}\equiv 4\pi/g_i^2$, for the three gauge coupling constants $g_1$, $g_2$, $g_3$.
The solid lines are the solution for the $m_D=4.6\times 10^8$ GeV and $m_Q = 2$ TeV case.
The dashed lines represent the gauge couplings of the Standard Model.
}
\end{figure}

\section{\label{sec:nonSUSY}Nonsupersymmetric scenario}

Apart from the observationally inconsistent prediction of the fermion mass relations \eqref{eqn:gutmassrel}, the Georgi-Glashow $SU(5)$ model presents three significant challenges: the three gauge coupling constants fail to unify at a single scale, the proton decay rate exceeds the experimental bounds, and substantial radiative corrections arise for light scalar particles (the hierarchy problem).
It is widely acknowledged that these challenges are effectively addressed by supersymmetry; 
in the next section, we will delve into the supersymmetric version of the $SU(5)$ model.
Here in this section, our focus will be on the nonsupersymmetric scenario, wherein at least the initial two aforementioned problems may be resolved through a minimally extended version of the original model.

\begin{table}[t]
\caption{\label{tab:table1}%
Matter contents added to the Standard Model in the nonsupersymmetric $SU(5)$ GUT scenario.
The representations in $SU(3)_c$, $SU(2)_L$, along with their corresponding $U(1)_Y$ charges and masses, are provided.
}
\begin{ruledtabular}
\begin{tabular}{c|cccl}
        field & $SU(3)_c$ & $SU(2)_L$ & $U(1)_Y$ & mass \\
        \hline \\
        $d'$ & $\bm{3}$ & $\bm{1}$ & $\sixth$ & \multirow{2}{*}{$\Big\}\; m_D=4.6\times 10^8$ GeV}\\
        $\overline{d}'$ & $\bm{3}^*$ & $\bm{1}$ & $-\sixth$ & \\
        $q'$ & $\bm{3}$ & $\bm{2}$ & $-\third$ & \multirow{2}{*}{$\Big\}\;m_Q=$ 2 TeV}\\
        $\overline{q}'$ & $\bm{3}^*$ & $\bm{2}$ & $\third$ &
\end{tabular}
\end{ruledtabular}
\end{table}

\subsection{\label{sec:nonSUSY_model}$SU(5)$ GUT without supersymmetry}%

Below the GUT scale, the Georgi-Glashow $SU(5)$ model has the same matter contents as the Standard Model and the three gauge coupling constants are known not to unify at one scale.
In order to achieve unification without supersymmetry, we consider introducing four fermion fields as listed in Table \ref{tab:table1}, in addition to the matter contents of the Standard Model (namely those of the Georgi-Glashow $SU(5)$ model below the GUT scale).
The fields $d'$, $\overline{d}'$, having the Dirac mass $m_D$, form a vectorlike pair and leave the anomaly cancellation intact.
They are split multiplets included in vectorlike $SU(5)$ fermions $\bm{5}+\bm{\overline 5}$ \cite{Frampton:1983sh,Amaldi:1991zx}.
Likewise, $q'$ and $\overline{q}'$ are split multiplets arising from a vectorlike pair $\bm{10}+\bm{\overline{10}}$. 
Thus we consider an $SU(5)$ GUT model with extra $\bm{5}+\bm{\overline 5}$ and $\bm{10}+\bm{\overline{10}}$ multiplets added to the Georgi-Glashow model.

The running of the gauge coupling constants, 
$g_3$ of $SU(3)_c$, $g_2$ of $SU(2)_L$ and $g_1=\sqrt{5/3}\, g_Y$ of $U(1)_Y$, is modified by the vectorlike fermions\footnote{
These vectorlike fermions are assumed to have no Yukawa coupling.
If their Yukawa coupling is nonzero, the Standard Model fermion masses can be affected through a seesaw-type mixing.
This effect has been discussed in \cite{Babu:2012pb,Dorsner:2014wva} as a potential resolution of the fermion mass problem.
In our case of $m_D=4.6\times 10^8$ GeV and $m_Q=2$ TeV, such corrections are negligibly small as long as the Yukawa coupling is perturbative.
}.
The renormalisation group equations for the gauge couplings are (at one-loop)
${dg_i}/{d\ln\mu}={g_i^3 b_i}/{16\pi^2}$,
with beta functions
\begin{align}
  b_i = (b_1,b_2,b_3)=&\; b_i^{\rm SM}+\Delta b_i^Q+\Delta b_i^D,\\
  b_i^{\rm SM} =& \left(\frac{41}{10},\; -\frac{19}{6},\; -7\right),\nn\\
  \Delta b_i^Q =& \left(\frac{2}{15},\; 2,\; \frac 43\right) \text{for } \mu>m_Q,\; 0 \text{ otherwise},\nn\\
  \Delta b_i^D =& \left(\frac{4}{15},\; 0,\; \frac{2}{3}\right) \text{for } \mu>m_D,\; 0 \text{ otherwise}.\nn
\end{align}  
%
%
When the vectorlike fermions have masses $m_D=4.6\times 10^8$ GeV and $m_Q = 2$ TeV, as listed in Table~\ref{tab:table1},  the three gauge couplings are found to unify\footnote{
We used the boundary conditions of \cite{Buttazzo:2013uya}.
}
at $M_U=1.13\times 10^{16}$ GeV.
Fig.~\ref{fig:NonSUSY_gRG} shows the flow of the gauge couplings in this case.
This unification scale is sufficiently high so that this scenario is consistent with the present proton decay bounds \cite{Super-Kamiokande:2016exg,Haba:2018vvu}.
We use this as our benchmark scenario of nonsupersymmetric $SU(5)$ GUT.

\begin{figure}[t]
\includegraphics[width=95mm]{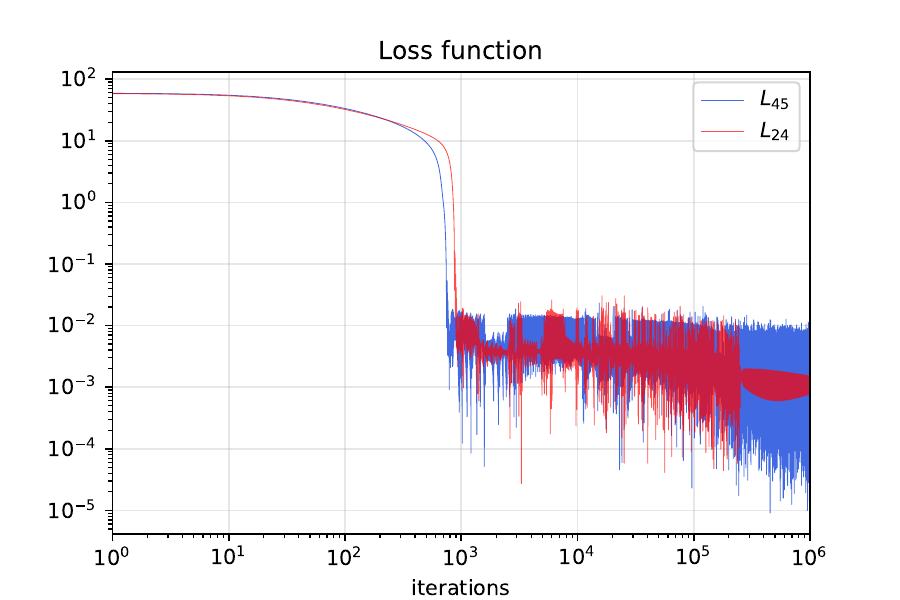}%
\caption
{\label{fig:LossEvo}
A sample of the loss function evolution for the 45-Higgs model (blue) and the 24-Higgs model (red), in the nonsupersymmetric scenario of Table~\ref{tab:table1}.
We selected $x_0=0$ and employed the identical set of 10 initial parameters for both the 45-Higgs model and the 24-Higgs model.
We utilised the Adam algorithm \cite{Kingma:2014vow} for optimisation and conducted iteration up to $N_{\rm iter}=10^6$ steps.
}
\end{figure}

The Yukawa couplings at the GUT scale are determined by the fermion masses at low energies through the renormalisation group equations.
Denoting the Yukawa matrices collectively as $y^f$, $f=\{u, d, e\}$ and introducing $S_f\equiv (y^f)^\dag y^f$ for convenience, the renormalisation group equations for the Yukawa couplings are written
\begin{align}\label{eqn:YukawaRGE}
	16\pi^2\frac{dS_f}{d\ln\mu}=\beta_fS_f+S_f\beta_f,
\end{align}
with beta functions
%
\begin{align}
	&\beta_u =\frac 32 S_u-\frac 32 S_d\crcr
	&\;+\left\{3\Tr S_u+3\Tr S_d+\Tr S_e-\frac{17}{20}g_1^2-\frac 94 g_2^2-8 g_3^2\right\}{\mathbb{1}},\crcr
	&\beta_d =\frac 32 S_d-\frac 32 S_u\crcr
	&\;+\left\{3\Tr S_u+3\Tr S_d+\Tr S_e-\frac{1}{4}g_1^2-\frac 94 g_2^2-8g_3^2\right\}{\mathbb{1}},\nonumber\\
	&\beta_e =\frac 32 S_e+\left\{3\Tr S_u+3\Tr S_d+\Tr S_e-\frac{9}{4}g_1^2-\frac 94 g_2^2\right\}{\mathbb{1}}.\nn\\
\end{align}
Here ${\mathbb{1}}$ is the $3\times 3$ unit matrix.
The boundary conditions at the low energy are given by
\begin{align}
	S_u =& {\rm diag}\left(\frac{2m_u^2}{v_h^2}, \frac{2m_c^2}{v_h^2}, \frac{2m_t^2}{v_h^2}\right),\nonumber\\
	S_d =& V_{\rm CKM}{\rm diag}\left(\frac{2m_d^2}{v_h^2}, \frac{2m_s^2}{v_h^2}, \frac{2m_b^2}{v_h^2}\right)V_{\rm CKM}^\dag,\nonumber\\
	S_e =& {\rm diag}\left(\frac{2m_e^2}{v_h^2}, \frac{2m_\mu^2}{v_h^2}, \frac{2m_\tau^2}{v_h^2}\right),
\end{align}
where $v_h = 246$ GeV.
The fermion masses and the CKM matrix at $\mu=M_Z$ are those given by \eqref{eqn:fmasses}.

\begin{figure*}
\includegraphics[width=85mm]{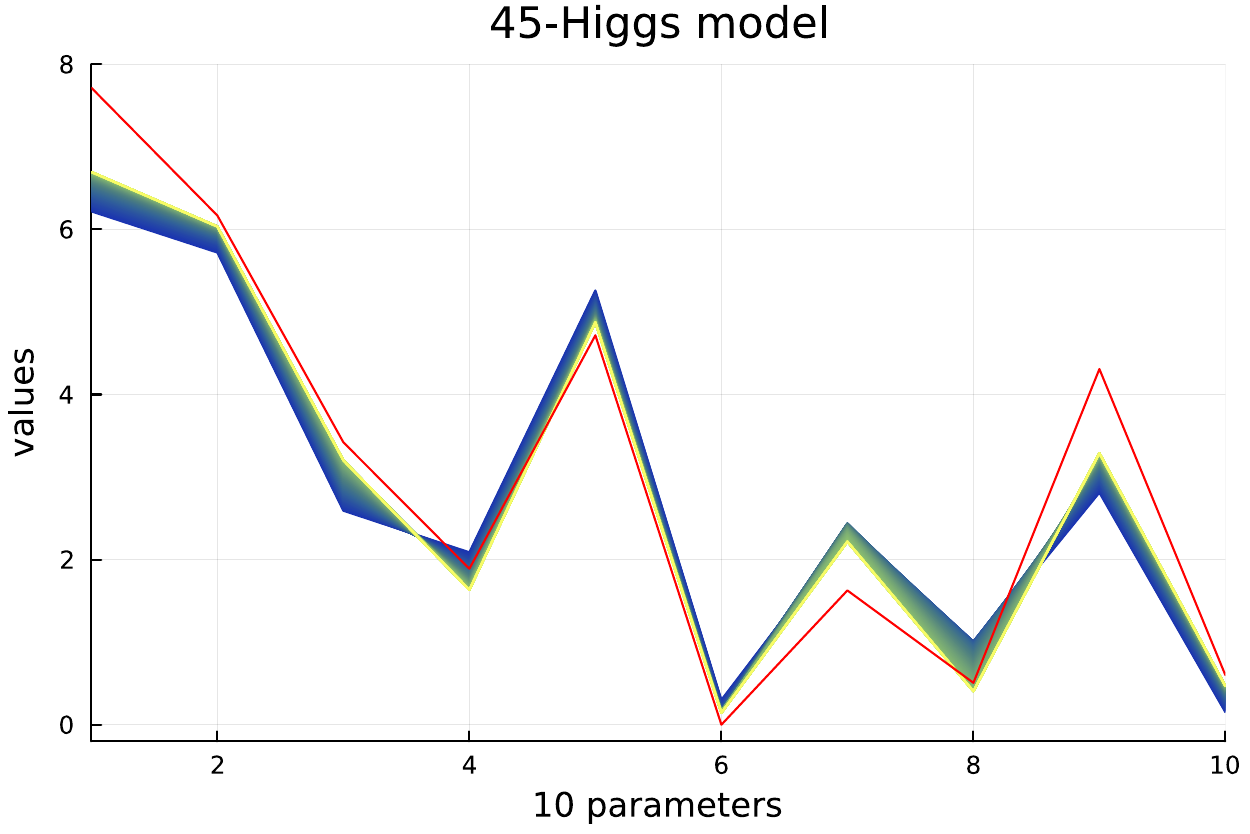}%
\includegraphics[width=85mm]{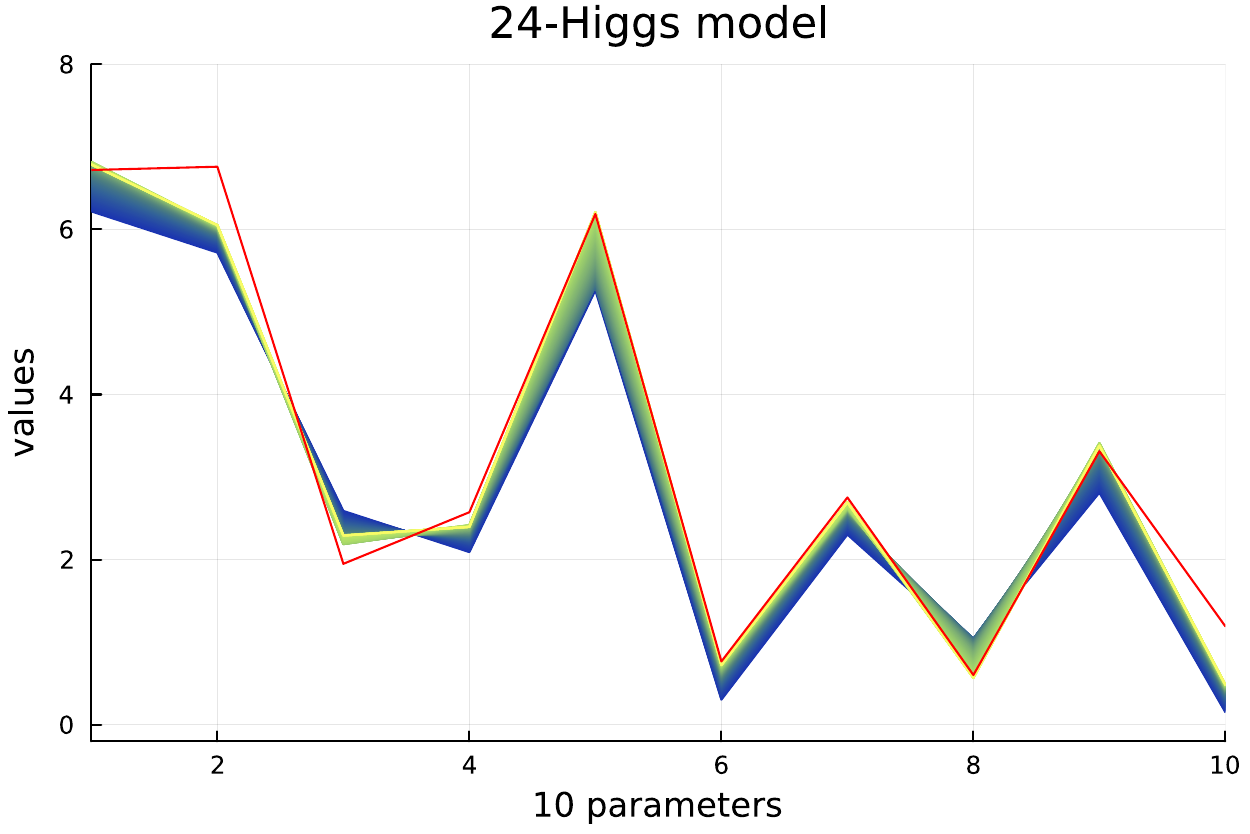}%
\caption
{\label{fig:ParamEvo}
Evolution of the ten parameters, $x_i$, $i=1,2,…,10$, in the same sample of optimisation processes shown in Fig.~\ref{fig:LossEvo}, for the 45-Higgs model (left panel) and the 24-Higgs model (right).
The darkest blue represents the initial random values of the parameters ($N_{\rm iter}=0$), which are chosen to be identical for the two models. The brightest yellow represents the parameter configuration at $N_{\rm iter}=1,000$, with an interval of 10 steps in between.
The red lines represent the configurations after $N_{\rm iter}=10^6$ steps.
The parameters of the two models begin with the same initial configurations and are observed to be adjusted to distinct optimised values. 
}
\end{figure*}

The Yukawa couplings (in the form of the eigenvalues of the mass matrices) and the CKM matrix at unification scale $\mu=M_U$ are found by integrating \eqref{eqn:YukawaRGE} up to the unification scale\footnote{
The equations are integrated from the top pole mass $M_t = 173.34$ GeV where the low-energy data at $M_Z$ instead of $M_t$ is used as boundary conditions; this serves as a suitable approximation.
}.
We find the fermion masses
\begin{widetext}
\begin{align}\label{eqn:nGUTmass}
m_u(M_U) &= 0.000496, &
m_c(M_U) &= 0.247, &
m_t(M_U) &= 75.9,\crcr
m_d(M_U) &= 0.00109, &
m_s(M_U) &= 0.0223, &
m_b(M_U) &= 1.01,\crcr
m_e(M_U) &= 0.000470, &
m_\mu(M_U) &= 0.0995, &
m_\tau(M_U) &= 1.69,  
\end{align}
in GeV and the CKM matrix parameters
\begin{align}\label{eqn:nGUTCKM}
%
	s_{12}(M_U) = 0.225,\quad
	s_{23}(M_U) = 0.0467,\quad
	s_{13}(M_U) = 0.00406,\quad
	\delta_{CP}(M_U) = 1.24.
\end{align}
\end{widetext}
We use these values in the following numerical analysis as input data for the nonsupersymmetric GUT scenario.

\subsection{\label{sec:nonSUSY_results}Good parameter values: numerical method}%

Under the assumption of this particular grand unification scenario, the fermion mass parameters and the CKM matrix at the GUT scale are determined as specified in \eqref{eqn:nGUTmass} and \eqref{eqn:nGUTCKM}.
The remaining unfixed parameters of the model are $x_0,\cdots,x_{10}$, as defined in equation \eqref{eqn:MdParam}.
Among these parameters, $x_0$ can assume one of the three discrete values: 0, $2\pi/3$, or $4\pi/3$, due to the constraints imposed by the vanishing strong CP phase.
The remaining parameters, $x_1,\cdots,x_{10}$, are phase parameters and can take continuous values within the range $0\leq x_i < 2\pi$.
The objective of our analysis is to identify within the parameter space the distribution of parameter values that are deemed beautiful.  
For the 45-Higgs and 24-Higgs models, and for each of the discrete parameter values of $x_0$, we carry out the analysis following three steps:
\begin{description}
  \item[Sampling] Generate initial values of the ten parameters $x_1, \ldots, x_{10}$ by randomly sampling from the uniform distribution $0 \leq x_i < 2\pi$.
  \item[Optimisation] Starting with the initial values, minimise the loss function \eqref{eqn:Ldef} using an optimisation scheme of machine learning.
  \item[Statistics] Repeat sampling and optimisation $N_{\rm samp}$ times and analyse the optimised loss function and the parameter values of the collected samples. 
\end{description}

For numerical optimisation, we employ the vanilla Adam algorithm \cite{Kingma:2014vow} with hyperparameters $\alpha=0.001$, $\beta_1=0.9$, $\beta_2=0.999$, and $\epsilon=10^{-8}$.
An example of optimisation process, with a single initial set of parameters, is presented in Figs.~\ref{fig:LossEvo} and~\ref{fig:ParamEvo}.
In this example, the discrete parameter is chosen to be $x_0=0$, and iteration is taken $N_{\rm iter}=10^6$ steps.
Fig.~\ref{fig:LossEvo} depicts the evolution of the loss function~\eqref{eqn:Ldef}, with the blue curve representing the 45-Higgs model and the red curve representing the 24-Higgs model.
Fig.~\ref{fig:ParamEvo} illustrates the evolution of the 10 parameter values during optimisation.
The 45-Higgs model (left) and the 24-Higgs model (right) commence with the same initial configuration (dark blue), but they are observed to converge to distinct optimised states (indicated by red lines).
Furthermore, this example demonstrates that the excursion of the parameter values is relatively small ($\lesssim 1$), suggesting that the loss function possesses numerous local minima.
Consequently, it is imperative to commence with a multitude of distinct initial configurations in order to explore the parameter space of the model.

\begin{figure*}[t]
\includegraphics[width=60mm]{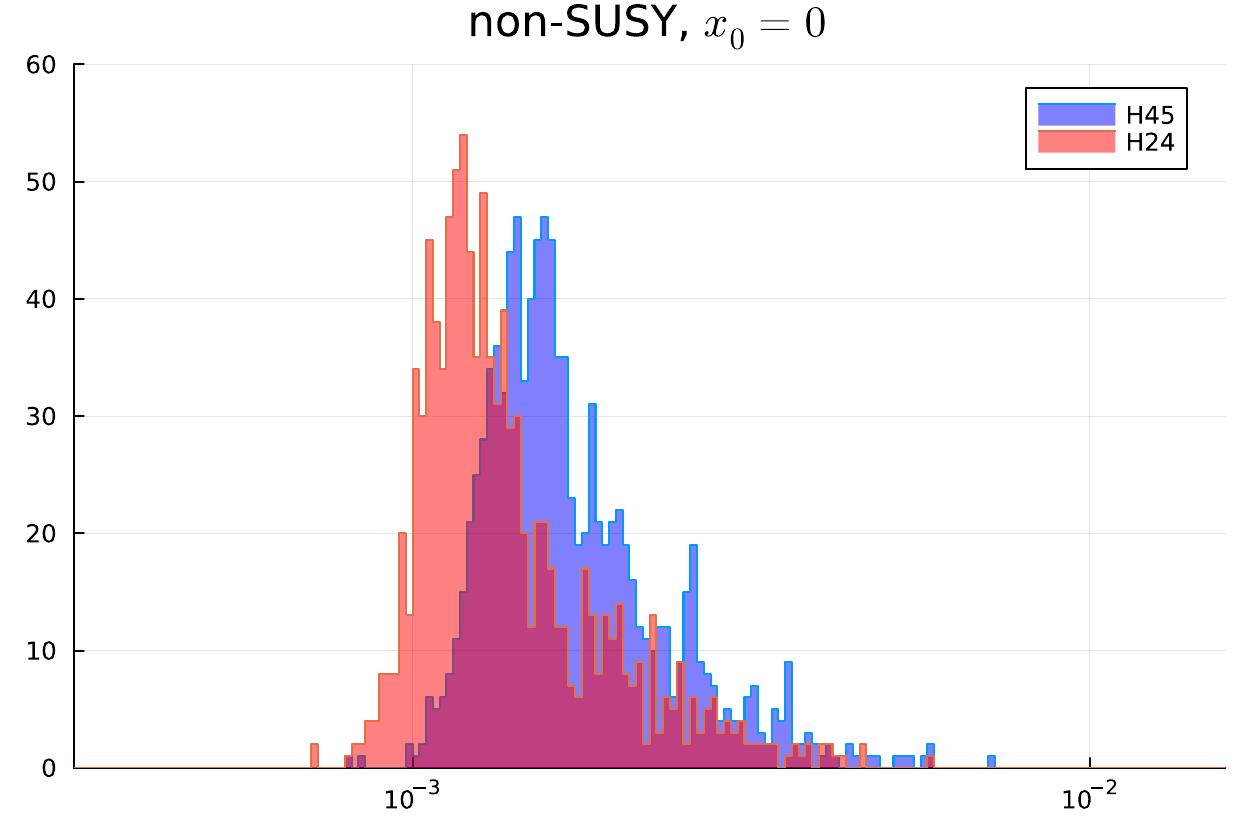}%
\includegraphics[width=60mm]{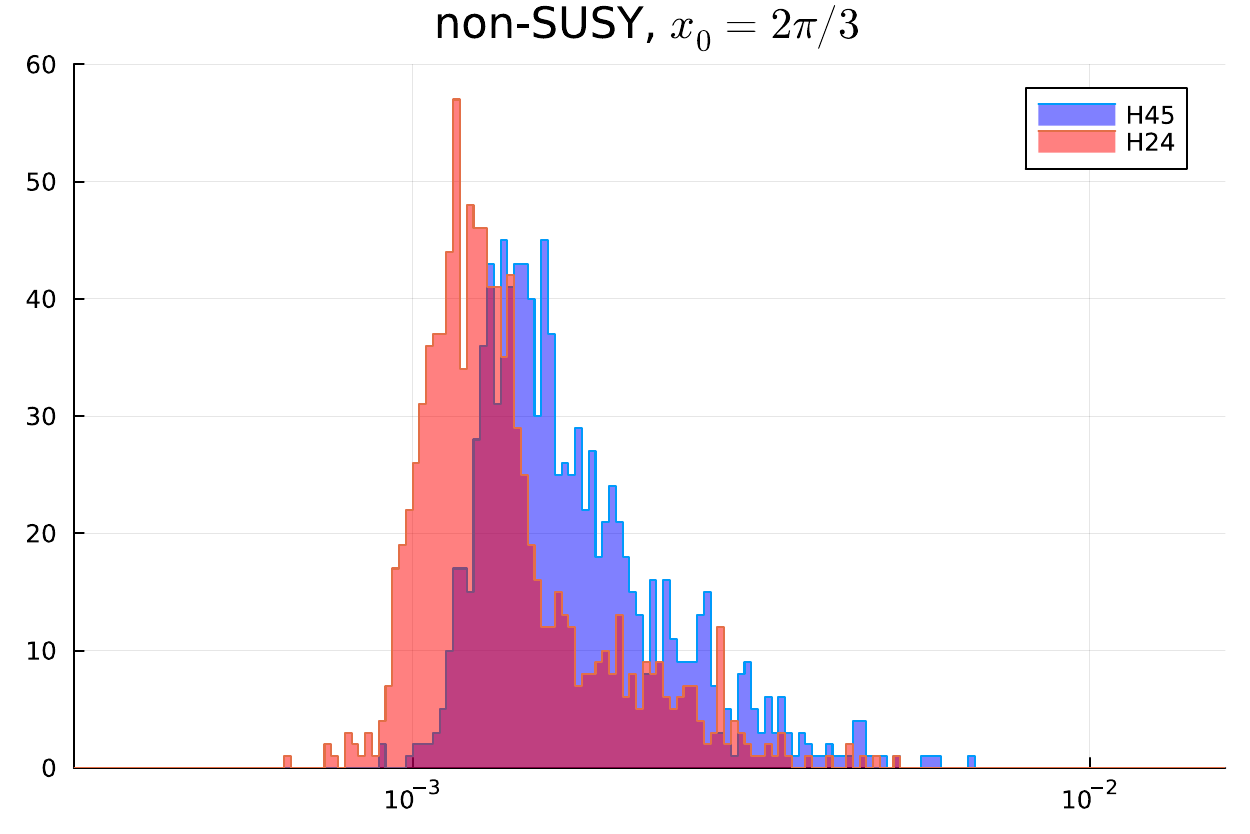}%
\includegraphics[width=60mm]{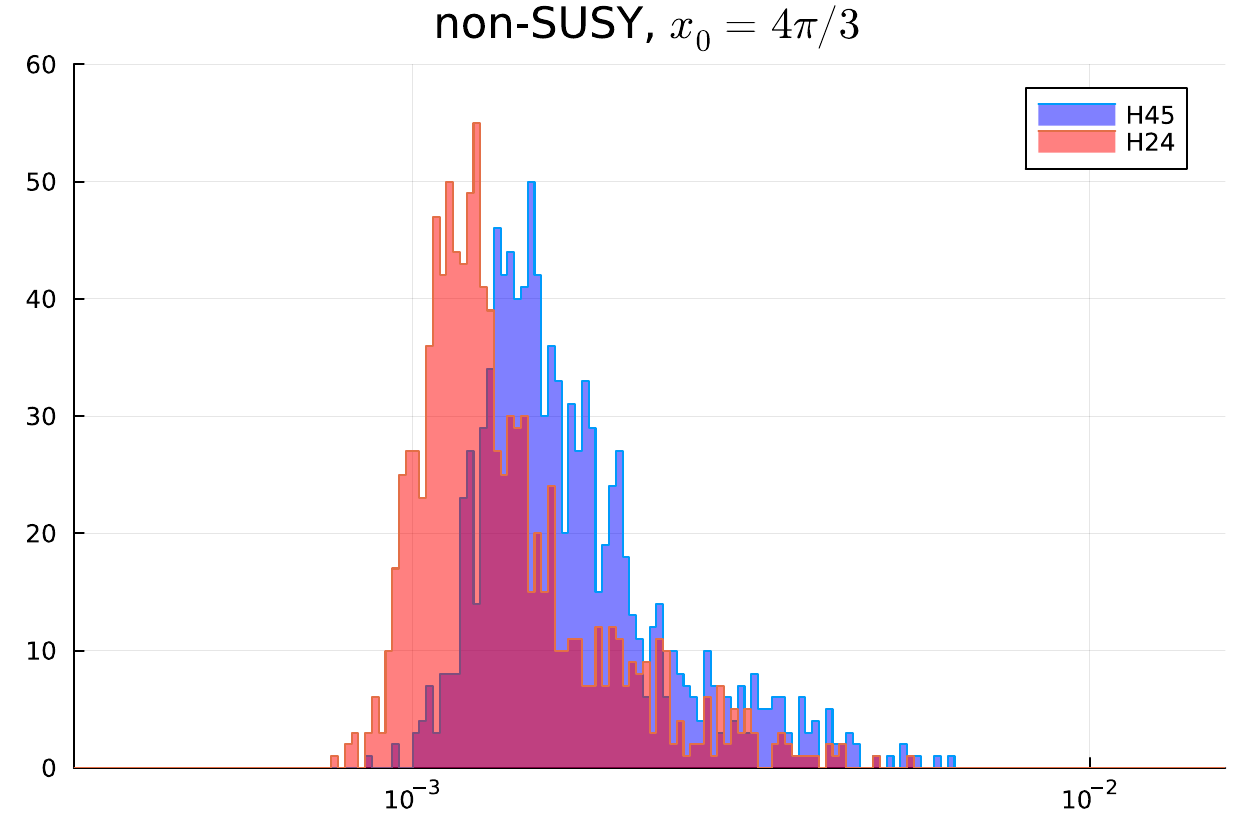}%
\caption
{\label{fig:nLhist}
Distribution of optimised loss function values for $N_{\rm samp} = 1024$ samples, in the nonsupersymmetric $SU(5)$ GUT scenario.
The left, middle, and right panels present the results for $x_0=0$, $2\pi/3$, and $4\pi/3$, respectively.
The 45-Higgs model (H45) is depicted in blue and the 24-Higgs model (H24) is shown in red.
An optimised value is determined by averaging over the final 100 steps of the total $N_{\rm iter}=10^6$ iterations, thereby mitigating the influence of fluctuations (refer to Fig.~\ref{fig:LossEvo}).
In all three cases, the loss function of the 24-Higgs model exhibits a distribution of smaller values compared to that of the 45-Higgs model. 
}
\end{figure*}

\subsection{\label{sec:nonSUSY_summary}45-Higgs model vs. 24-Higgs model}%

We collected 1024 samples of numerical optimisation with randomly chosen initial configurations, each for the 45-Higgs and 24-Higgs models and for the discrete parameter values $x_0=0$, $2\pi/3$ and $4\pi/3$. 
The distribution of the 1024 minimised values of the loss function, defined in \eqref{eqn:Ldef}, for each case is shown in Fig.~\ref{fig:nLhist}.
Here, we evaluated the minimised loss function value by averaging over the final 100 steps (i.e. 999,901st-1,000,000th) of the total $N_{\rm iter}=10^6$ iterations, in order to mitigate the effects of fluctuations (see Fig.~\ref{fig:LossEvo}). 
In all three cases of $x_0=0$, $2\pi/3$ and $4\pi/3$, the loss function of the 24-Higgs model, depicted in red, is seen to be distributed at smaller values than that of the 45-Higgs model (blue).
This indicates the tendency that the 24-Higgs model lies closer to the original Georgi-Glashow model than the 45-Higgs model does.
Thus, according to our definition of beauty, the 24-Higgs model is more beautiful than the 45-Higgs model.

\begin{figure*}
\includegraphics[width=60mm]{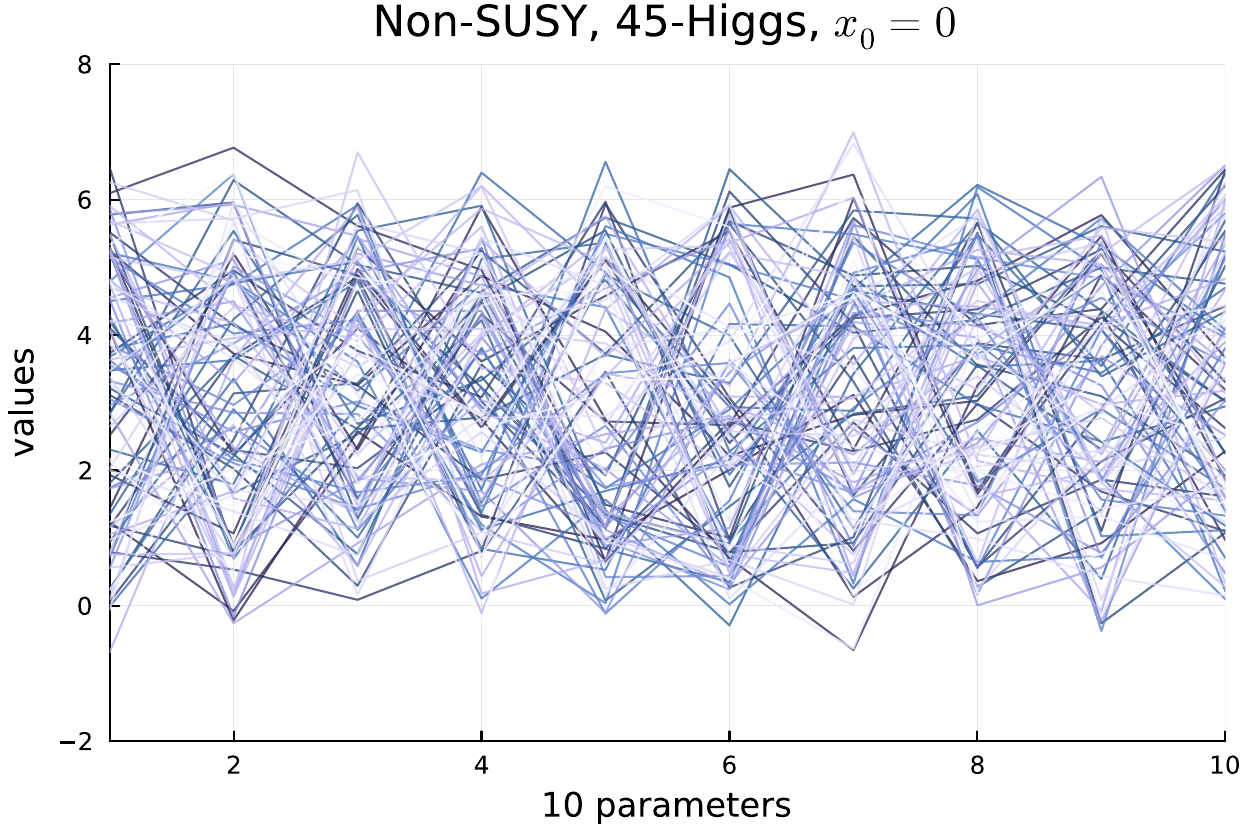}%
\includegraphics[width=60mm]{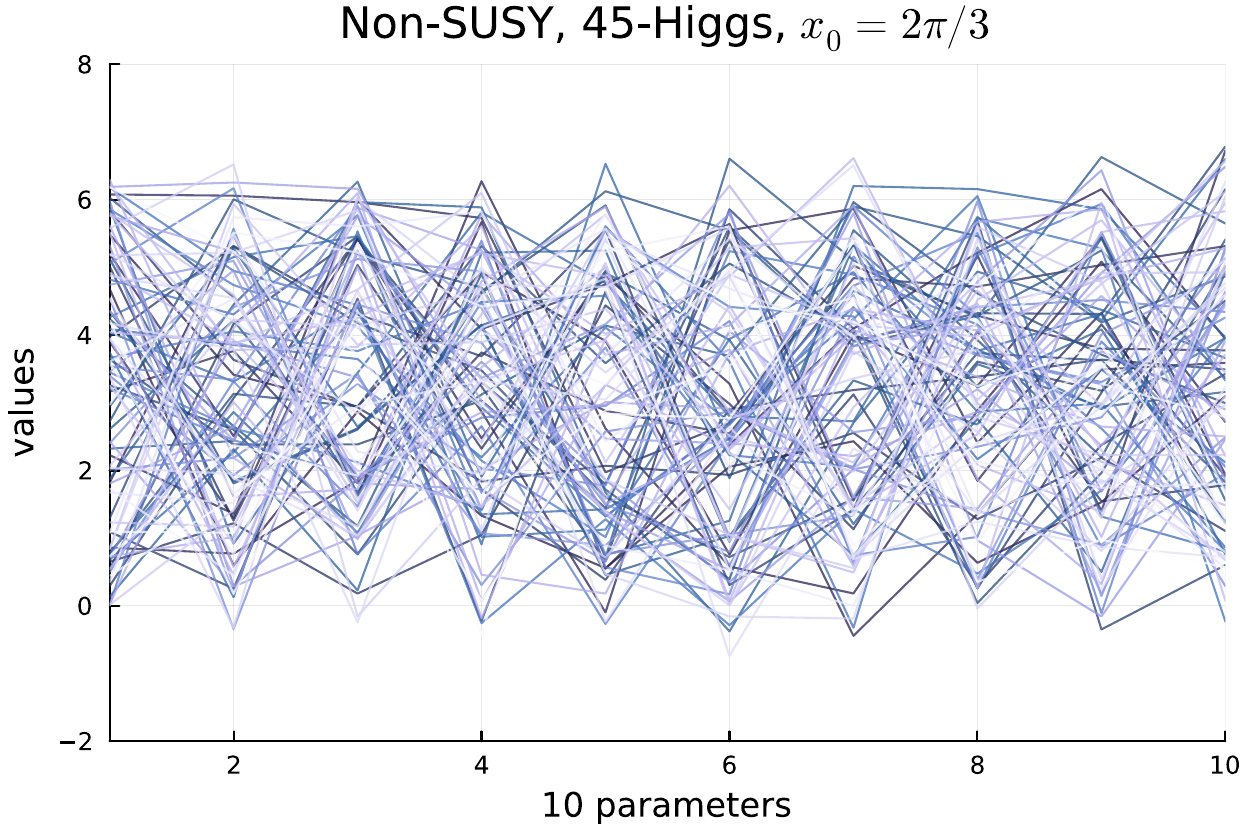}%
\includegraphics[width=60mm]{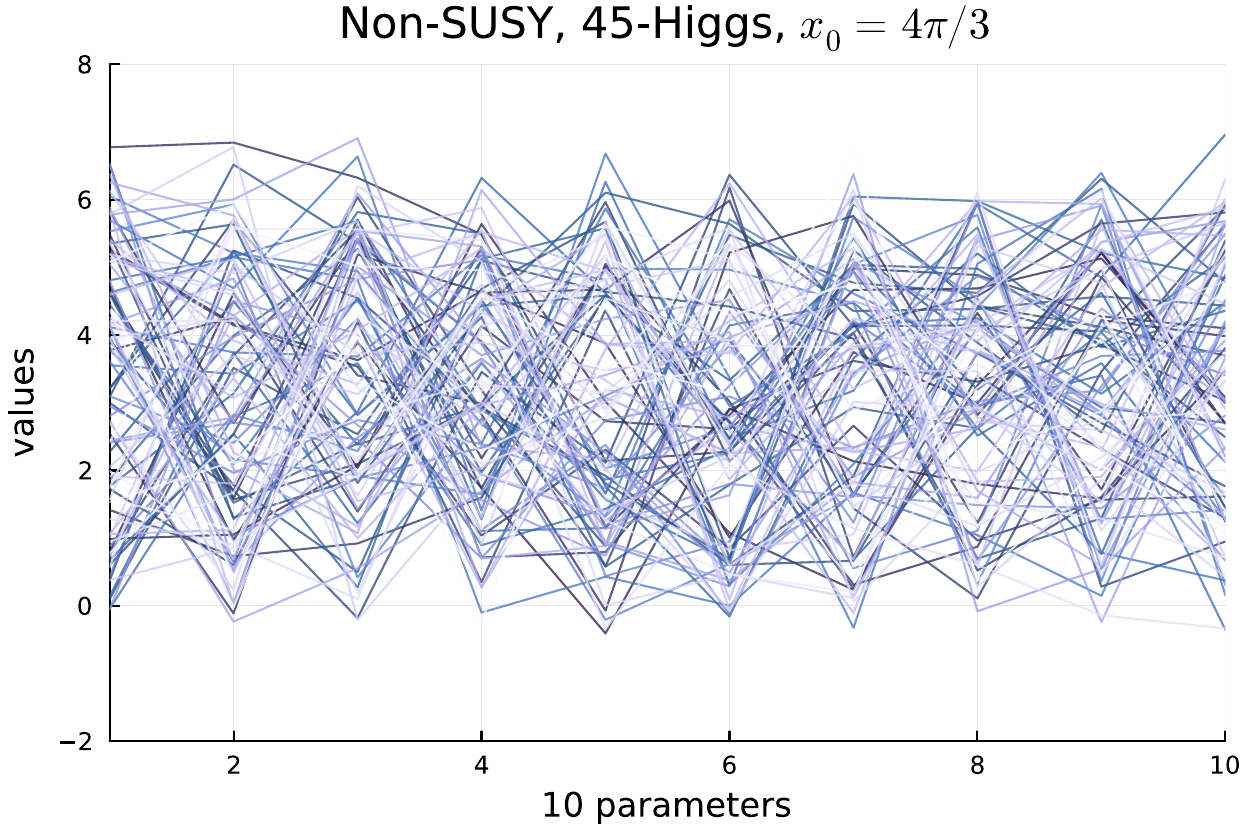}\\
\includegraphics[width=60mm]{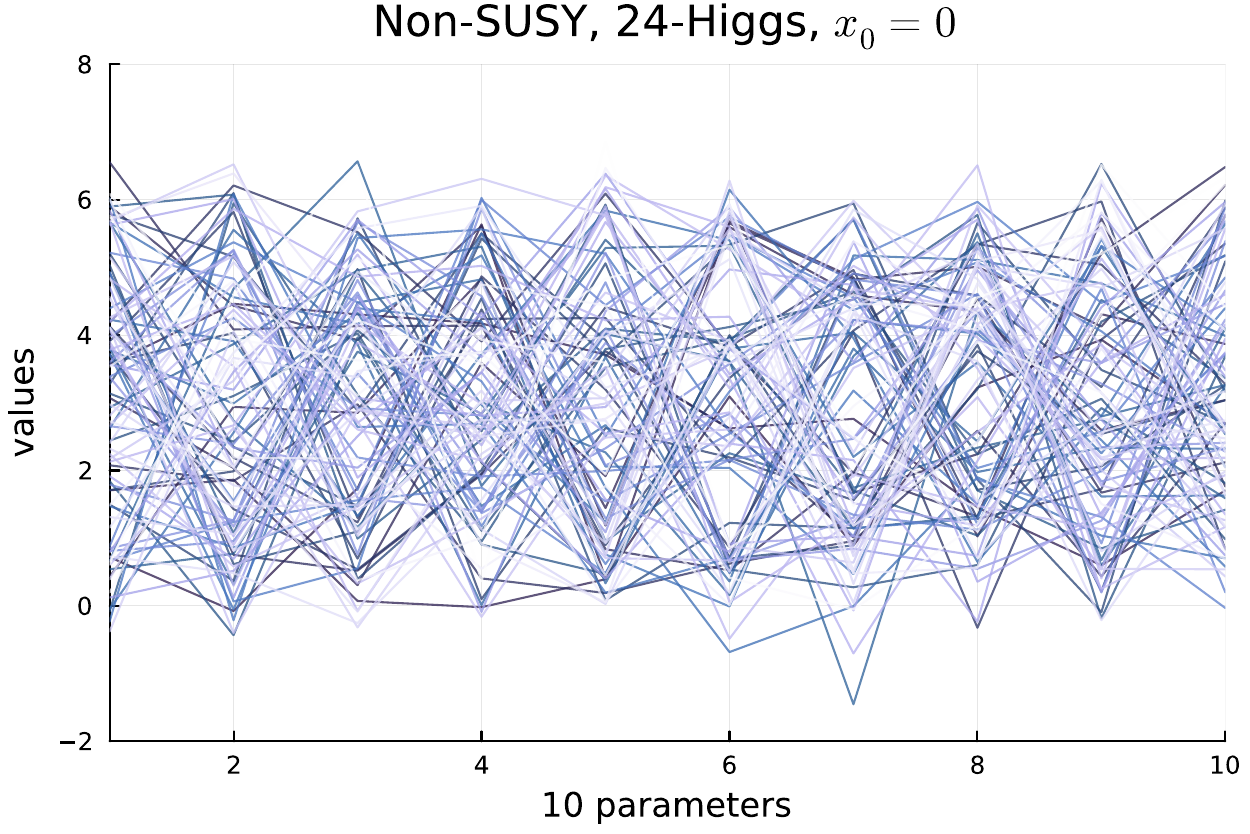}%
\includegraphics[width=60mm]{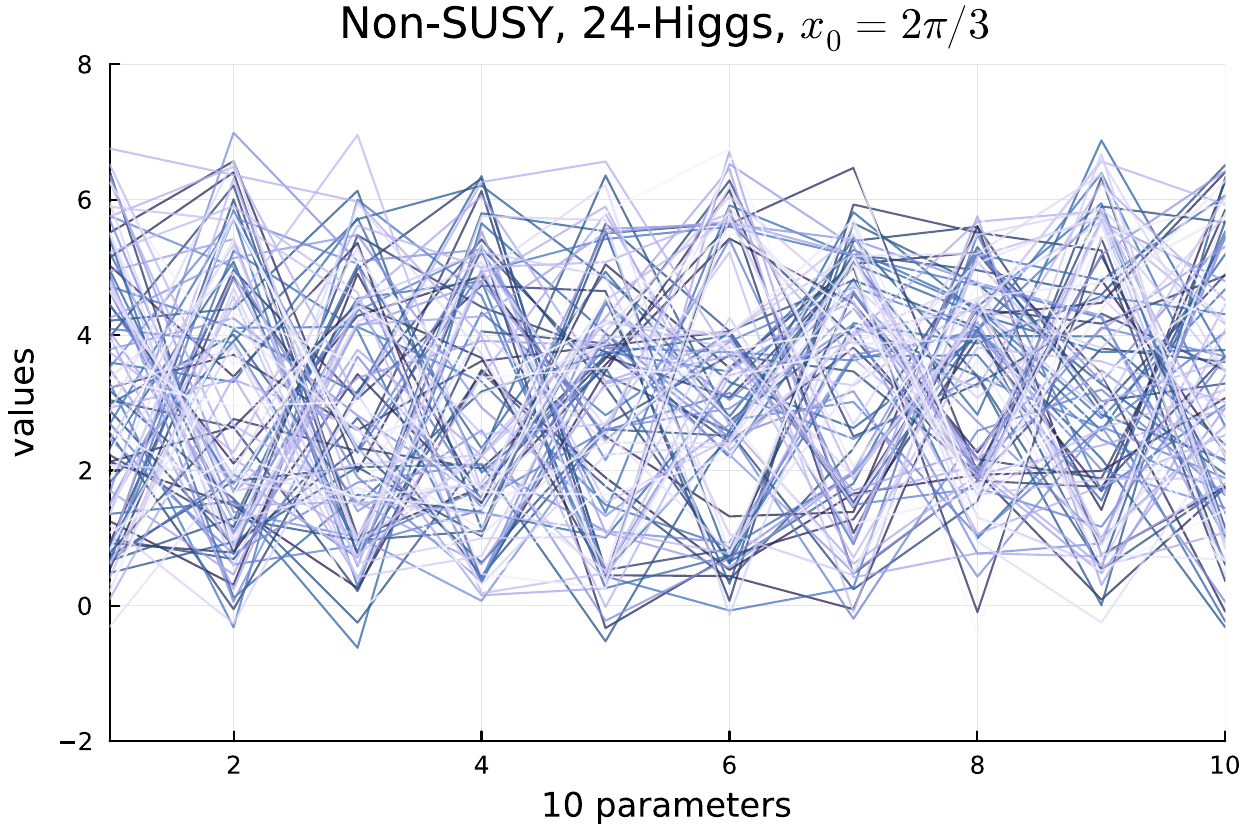}%
\includegraphics[width=60mm]{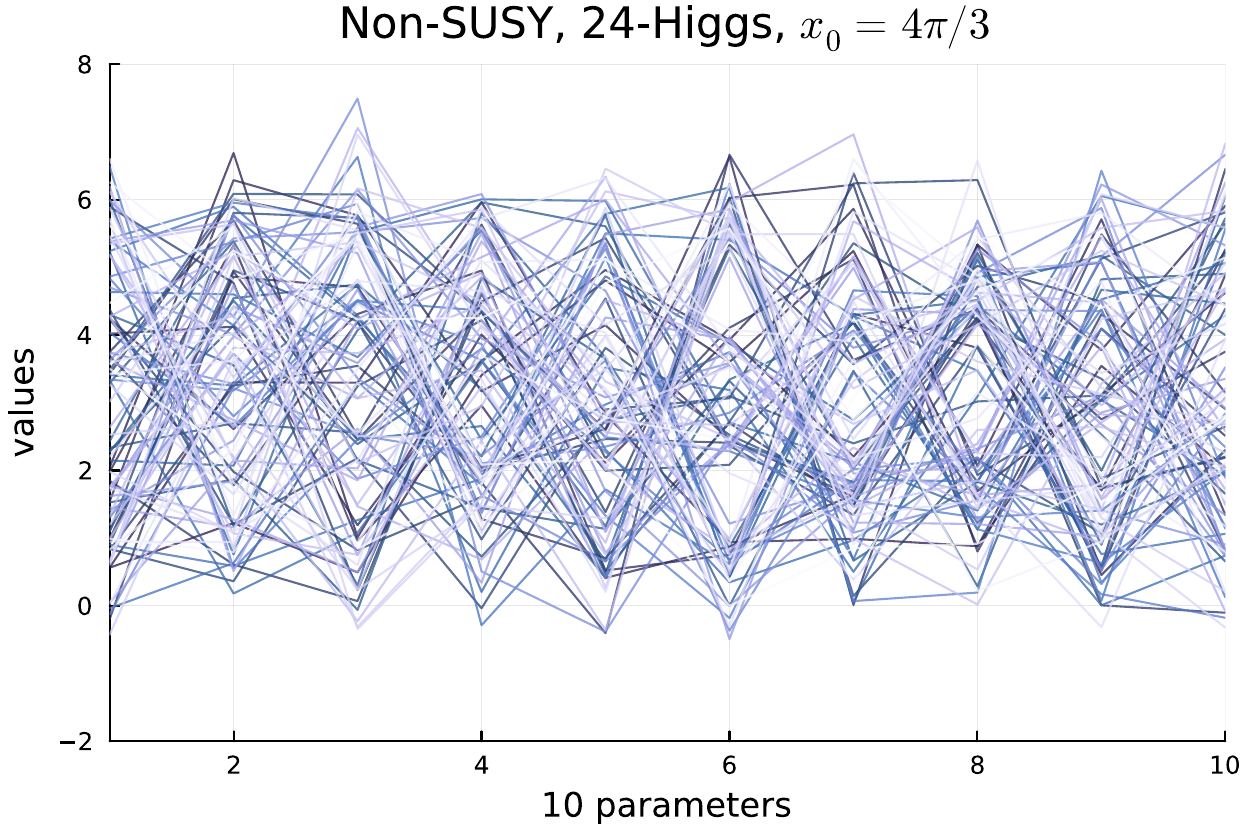}%
\caption
{\label{fig:nxVals}
Optimised configurations of ten parameters $x_1,\ldots,x_{10}$ in the nonsupersymmetric $SU(5)$ GUT scenario.
The upper (lower) panels present the results for the 45-Higgs (24-Higgs) model. The cases where $x_0=0$, $2\pi/3$, and $4\pi/3$ are depicted on the left, middle, and right panels, respectively.
Each panel displays 100 samples representing the 100 smallest values of the loss function out of the 1024 available samples. 
Darker lines correspond to the smaller loss function values.
}
\end{figure*}

As illustrated in Fig.~\ref{fig:ParamEvo}, the parameters $x_1,\ldots,x_{10}$ are adjusted to their optimal values, as indicated by the red lines, to minimise the loss function.
Fig.~\ref{fig:nxVals} presents the optimised configurations of parameters $x_1,\ldots,x_{10}$ for the 45-Higgs and 24-Higgs models, for values of $x_0=0$, $2\pi/3$ and $4\pi/3$.
Each panel displays only 100 samples out of 1024, representing the 100 smallest values of the minimised loss function. Darker (lighter) blue colours correspond to smaller (larger) loss function values.
A smaller loss function indicates a more beautiful parameter configuration.
Consequently, regions in darker blue indicate preferred parameter values, while blank regions are disfavoured.

The three values of $x_0$ all give similar results. 
Thus our criterion of beauty does not favour or disfavour a particular value of $x_0$ that avoids the strong CP problem.

\section{\label{sec:SUSY}Supersymmetric scenario}

\subsection{\label{sec:SUSY_model}Supersymmetric $SU(5)$ GUT}%

The supersymmetric version of the minimal $SU(5)$ GUT consists of a $\bm{24}$ representation vector multiplet $V_{\bm{24}}$, one $\bm{24}$, one $\bm{5}$, one $\bm{\overline 5}$ and three copies of $\bm{10}$ and $\bm{\overline 5}$ representation chiral multiplets of $SU(5)$ that we denote by $H_{\bm{24}}$, $H_{\bm{5}}$, $H_{\bm{\overline 5}}$, $F^i_{\bm{10}}$ and $F^i_{\bm{\overline 5}}$ respectively.
The $SU(5)$ fermions $\Psi^i_{\bm{\overline 5}}$ of Eq.~\eqref{eqn:F5} and $\Psi^i_{\bm{10}}$ of Eq.~\eqref{eqn:F10} are the fermionic parts of $F^i_{\bm{\overline 5}}$ and $F^i_{\bm{10}}$.
The bosonic parts of the chiral multiplets $H_{\bm{24}}$ and $H_{\bm{5}}$ are the $\bm{24}$ and $\bm{5}$ representation Higgs fields represented by the same symbols.
While $H_{\bm{\overline 5}}$ in \eqref{eqn:YukawaL} was the complex conjugate of $H_{\bm{5}}$ in the nonsupersymmetric scenario, here $H_{\bm{\overline 5}}$ is (the bosonic part of) a separate chiral supermultiplet.
Consequently the $\bm{5}$ and $\bm{\overline 5}$ Higgs may acquire distinct vacuum expectation values
\begin{align}\label{eqn:sH5vev}
	\langle H_{\bm{5}}\rangle=[0, 0, 0, 0, \frac{v_u}{\sqrt 2}], \qquad
	\langle H_{\bm{\overline 5}}\rangle=[0, 0, 0, 0, \frac{v_d}{\sqrt 2}].
\end{align}
We assume their ratio to be fixed, $\tan\beta\equiv v_u/v_d = 10$, in the numerics.
The superpotential of the minimal supersymmetric $SU(5)$ model is given by
\begin{align}
	W_{\rm min}=&\frac{m_\Sigma}{2} \Tr\left( H_{\bm{24}} \right)^2
	+\frac{\lambda_\Sigma}{3} \Tr\left(H_{\bm{24}}\right)^3\crcr
	&+H_{\bm{\overline 5}} \left(m_H+\lambda_H H_{\bm{24}}\right)H_{\bm{5}}\crcr
	&-y^{5d}_{ij}[F^i_{\bm{\overline 5}}]^m[F^j_{\bm{10}}]_{mn}[H_{\bm{5}}]^n\crcr
	&-y^{5u}_{ij}\epsilon^{mnpqr}[F^i_{\bm{10}}]_{mn}[F^j_{\bm{10}}]_{pq}[H_{\bm{\overline 5}}]_r,
\end{align}
where $m_\Sigma$, $\lambda_\Sigma$, $m_H$ and $\lambda_H$ are real parameters.
The Yukawa part of the Lagrangian is
\begin{align}
	y^{5d}_{ij}\frac{v_d}{\sqrt 2}[\overline{\Psi^i_{\bm{\overline 5}}}]^m[\Psi^j_{\bm{10}}]_{m5}
	&+y^{5u}_{ij}\frac{v_u}{\sqrt 2}\epsilon^{mnpq5}[\overline{\Psi^i_{\bm{10}}}]_{mn}[\Psi^j_{\bm{10}}]_{pq}\nonumber\\
	&\qquad\qquad\qquad\qquad\qquad+{\rm h.c.},
\end{align}
and the mass matrices are
\begin{align}\label{eqn:sM5rel}
	M_u =4\frac{v_u}{\sqrt 2} y^{5u}_{ij},\;\;
	M_d = \frac{v_d}{2} y^{5d}_{ij},\;\;
	M_e = \frac{v_d}{2} y^{5d}_{ji} = M_d^T,
\end{align}
leading to the same GUT mass relations \eqref{eqn:gutmassrel}.

The supersymmetric version of the 45-Higgs model posits the presence of an additional chiral superfield, denoted as $H_{\bm{45}}$, along with its corresponding superpotential

\begin{align}
	W_{45}=&\;y^{45d}_{ij}[F^i_{\bm{\overline 5}}]^m[F^j_{\bm{10}}]_{np}[H_{\overline{\bm{45}}}]^{np}_m.
\end{align}
Assuming the vacuum expectation value \eqref{eqn:H45vev}, the mass matrices are modified from \eqref{eqn:sM5rel} as
\begin{align}\label{eqn:sM45rel}
	M_u =& 2\sqrt 2 v_u y^{5u}_{ij},\nonumber\\
	M_d =& \frac{v_d}{2} y^{5d}_{ij}+ \frac{v_{45}}{\sqrt 2}\, y^{45d}_{ij},\nonumber\\
	M_e =& \frac{v_d}{2} y^{5d}_{ji}-3\frac{v_{45}}{\sqrt 2}\, y^{45d}_{ji}.
\end{align}

The alternative, 24-Higgs approach incorporates contributions from a higher dimensional $SU(5)$ gauge singlet term 
$F_{\bm{\overline 5}}H_{\bm{24}}F_{\bm{10}}H_{\bm{\overline 5}}\subset W$ which gives rise to the term in the Lagrangian \eqref{eqn:H24lag}.
The GUT Higgs field $H_{\bm{24}}$ acquires a vacuum expectation value
$H_{\bm{24}}=(m_\Sigma/\lambda_\Sigma)\,{\rm diag}(2,2,2,-3,-3)$.
The expectation value of $H_{\bm{\overline 5}}$ is specified by \eqref{eqn:sH5vev}.
Then the fermion mass matrices in this case become
\begin{align}\label{eqn:sM24rel}
	M_u =& 2\sqrt 2 v_u y^{5u}_{ij},\nonumber\\
	M_d =& \frac{v_d}{2} y^{5d}_{ij}+ \frac{m_\Sigma}{m_{\rm P}\lambda_\Sigma}v_d\, y^{24d}_{ij},\nonumber\\
	M_e =& \frac{v_d}{2} y^{5d}_{ji}- \frac{3}{2}\frac{m_\Sigma}{m_{\rm P}\lambda_\Sigma}v_d\, y^{24d}_{ji}.
\end{align}

Thus the formulation of the fermion mass matrices parallels the nonsupersymmetric case. 
Introducing 
\begin{align}
	&M_5 \equiv  \frac{v_d}{2} y^{5d}_{ij},\quad
	M_{45} \equiv \frac{v_{45}}{\sqrt 2}\, y^{45d}_{ij},\crcr
	&M_{24} \equiv \frac{m_\Sigma}{2m_{\rm P}\lambda_\Sigma}v_d\, y^{24d}_{ij},
\end{align}
the modified GUT mass relations of the 45-Higgs model \eqref{eqn:sM45rel} and the 24-Higgs model \eqref{eqn:sM24rel} are written in the same forms as the nonsupersymmetric scenario, \eqref{eqn:MM45} and \eqref{eqn:MM24}.

\begin{figure*}
\includegraphics[width=60mm]{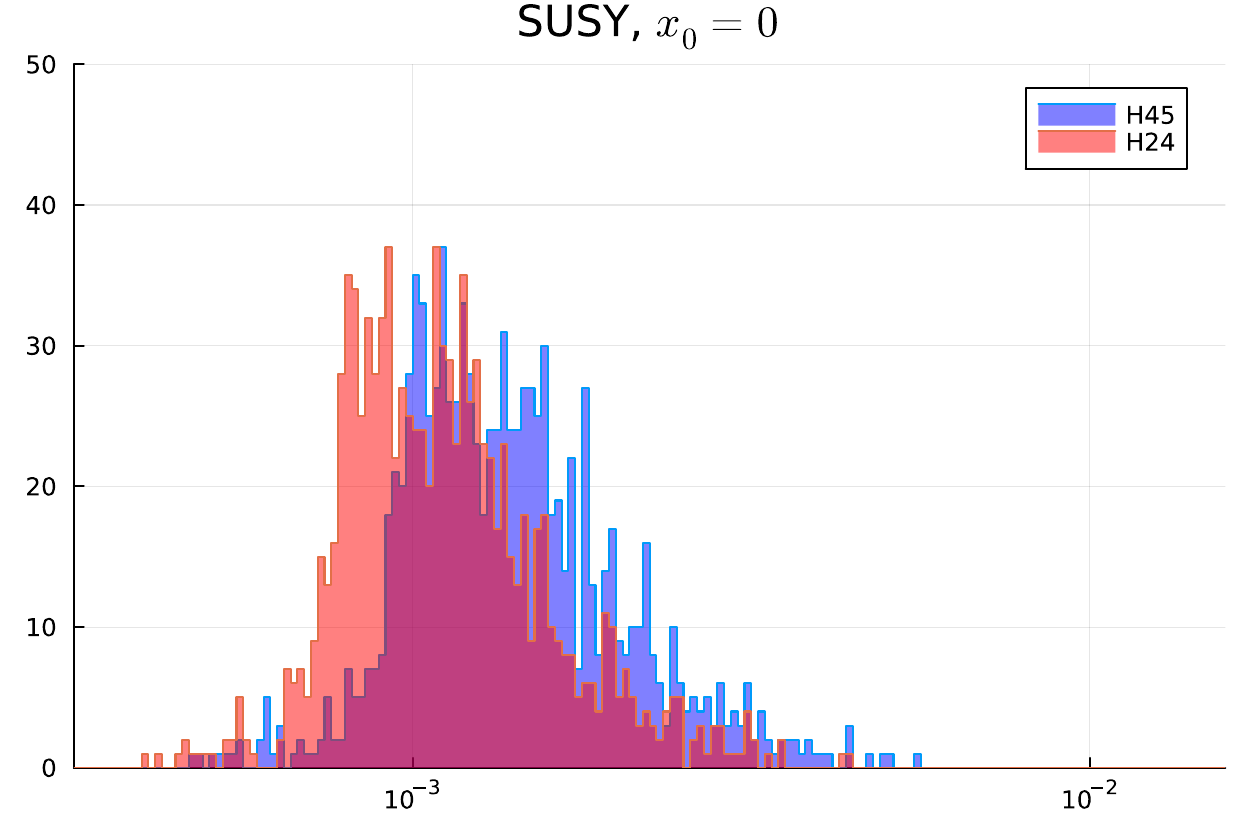}%
\includegraphics[width=60mm]{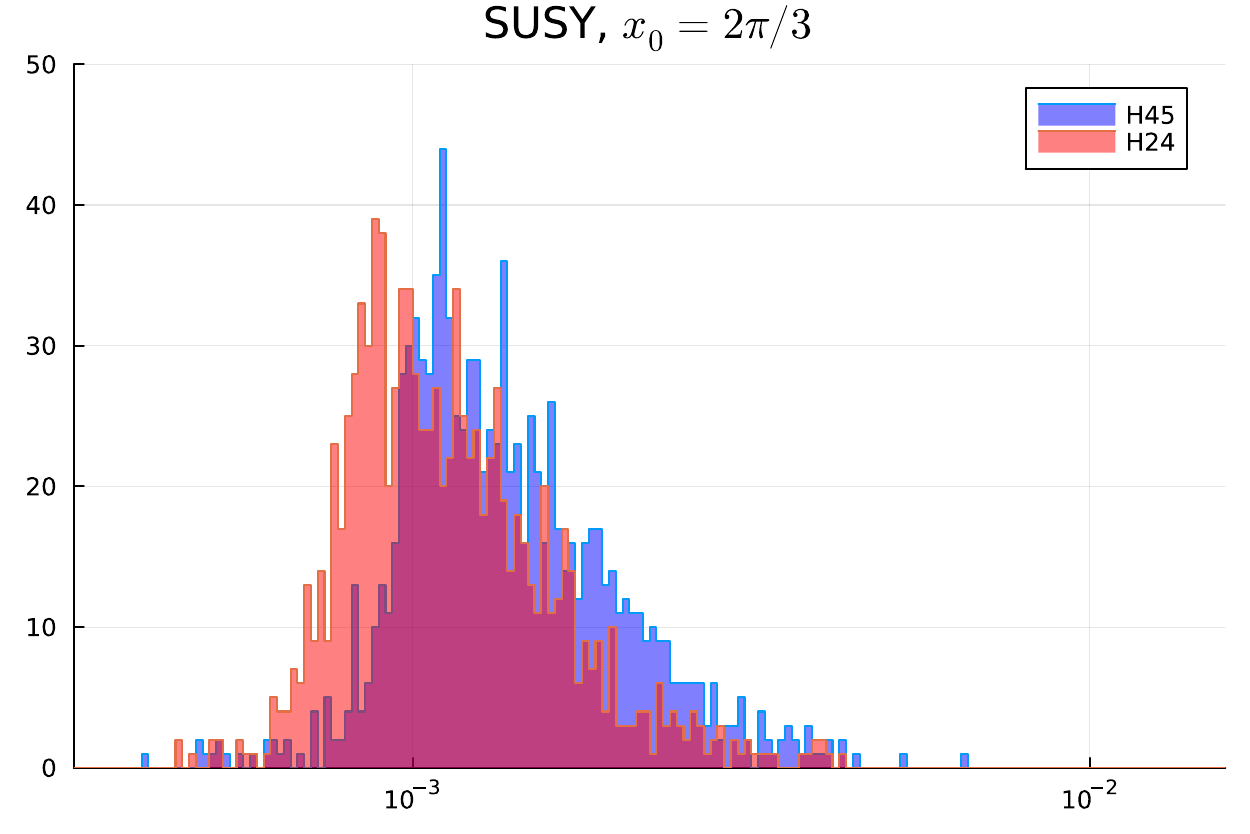}%
\includegraphics[width=60mm]{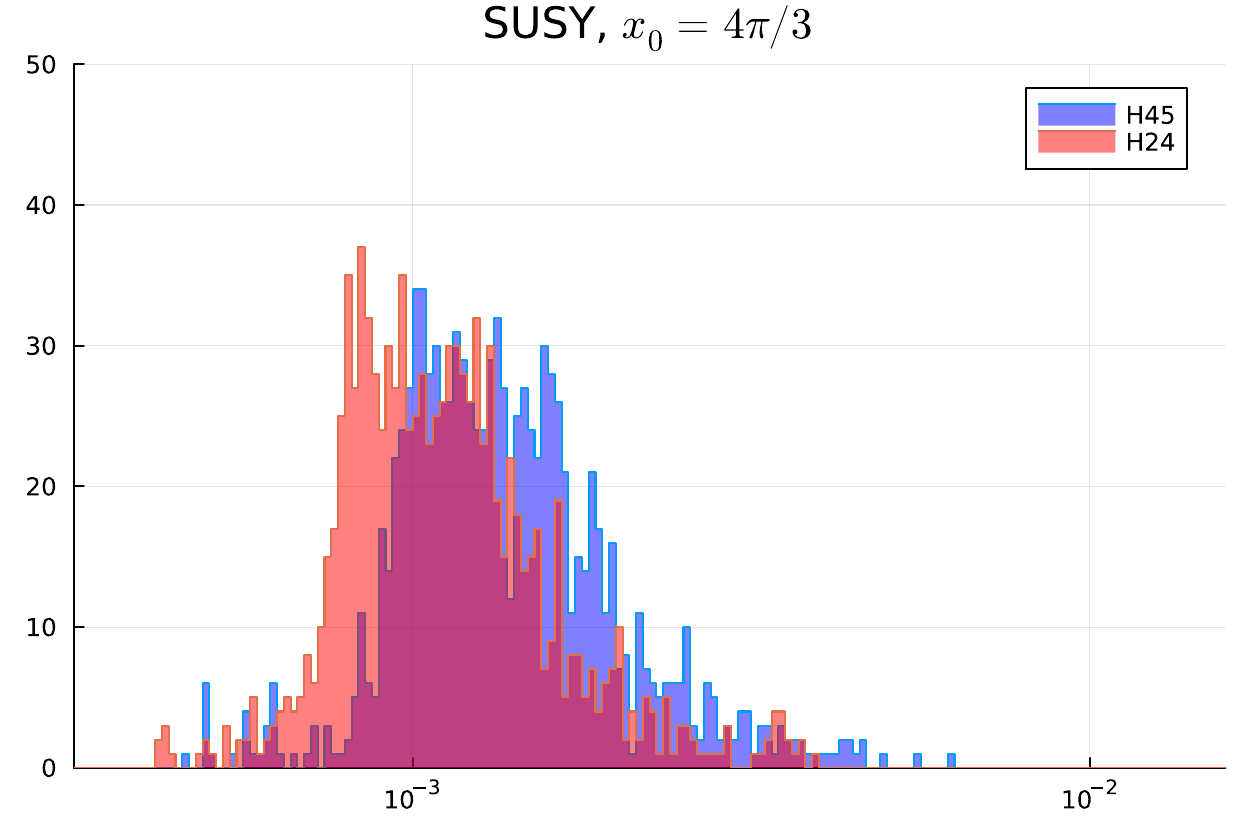}%
\caption
{\label{fig:sLhist}
Distribution of optimised loss function values for $N_{\rm samp} = 1024$ samples, in the supersymmetric $SU(5)$ GUT scenario.
The left, middle, and right panels present the results for $x_0=0$, $2\pi/3$, and $4\pi/3$, respectively.
The 45-Higgs model (H45) is depicted in blue and the 24-Higgs model (H24) is shown in red.
An optimised value is determined by averaging over the final 100 steps of the total $N_{\rm iter}=10^6$ iterations, thereby mitigating the influence of fluctuations (refer to Fig.~\ref{fig:LossEvo}).
In all three cases, the loss function of the 24-Higgs model exhibits a distribution of smaller values compared to that of the 45-Higgs model. 
}
\end{figure*}

Assuming low supersymmetry breaking scale $M_S\simeq 10$ TeV, we use the renormalisation group equations of the MSSM from $\mu=M_Z=91.2$ GeV\footnote{
This approximation is consistent in the parameter region where the Super-Kamiokande bounds \cite{ParticleDataGroup:2024cfk} on the proton decay through the 5-dimensional operator are satisfied.
We also neglected threshold corrections, which are discussed e.g. in \cite{Antusch:2008tf}.
}.
At one loop, the beta functions for the gauge couplings are 
$b_i\equiv (b_1, b_2, b_3)=(33/5,\, 1,\, -3)$
and the three couplings unify at $M_U\simeq 3\times 10^{16}$ GeV.
The renormalisation group equations for the Yukawa couplings \eqref{eqn:YukawaL} now have beta functions \cite{Castano:1993ri}
\begin{align}
	\beta_u=&3S_u+S_d+\left\{\Tr(3S_u)-\frac{13}{15}g_1^2-3g_2^2-\frac{16}{3}g_3^2\right\}{\mathbb{1}},\nonumber\\
	\beta_d=&3S_d+S_u+\left\{\Tr(3S_d+S_e)-\frac{7}{15}g_1^2-3g_2^2-\frac{16}{3}g_3^2\right\}{\mathbb{1}},\nonumber\\
	\beta_e=&3S_e+\left\{\Tr(3S_d+S_e)-\frac{9}{5}g_1^2-3g_2^2\right\}{\mathbb{1}}.
\end{align}
The boundary conditions for the Yukawa couplings at $\mu=M_Z$ are set as
\begin{align}
	S_u =& {\rm diag}\left(\frac{m_u^2}{v^2}, \frac{m_c^2}{v^2}, \frac{m_t^2}{v^2}\right)\left(1+\frac{1}{\tan^2\beta}\right),\nonumber\\
	S_d =& V_{\rm CKM}{\rm diag}\left(\frac{m_d^2}{v^2}, \frac{m_s^2}{v^2}, \frac{m_b^2}{v^2}\right)V_{\rm CKM}^\dag (1+\tan^2\beta),\nonumber
\end{align}\\
\begin{align}
	S_e =& {\rm diag}\left(\frac{m_e^2}{v^2}, \frac{m_\mu^2}{v^2}, \frac{m_\tau^2}{v^2}\right)\left(1+\frac{1}{\tan\beta}\right),
\end{align}
in the basis where $S_u$ and $S_e$ are diagonal.
Solving the renormalisation group equations the fermion mass parameters at the GUT scale $\mu=M_U$ are found to be
\begin{widetext}
\begin{align}\label{eqn:sGUTmass}
  	m_u(M_U) &= 0.000502, & m_c(M_U)   &= 0.251,  & m_t(M_U)    &= 97.7, \nonumber\\
	m_d(M_U) &= 0.000769, & m_s(M_U)   &= 0.0157, & m_b(M_U)    &= 0.922, \nonumber\\
	m_e(M_U) &= 0.000324, & m_\mu(M_U) &= 0.0685, & m_\tau(M_U) &= 1.17, 
\end{align}
in GeV.
The CKM matrix parameters are
\begin{align}\label{eqn:sGUTCKM}
%
	s_{12}(M_U) = 0.225,\quad
	s_{23}(M_U) = 0.0363,\quad
	s_{13}(M_U) = 0.00315,\quad
	\delta_{CP}(M_U) = 1.24.
\end{align}
\end{widetext}

\begin{figure*}
\includegraphics[width=60mm]{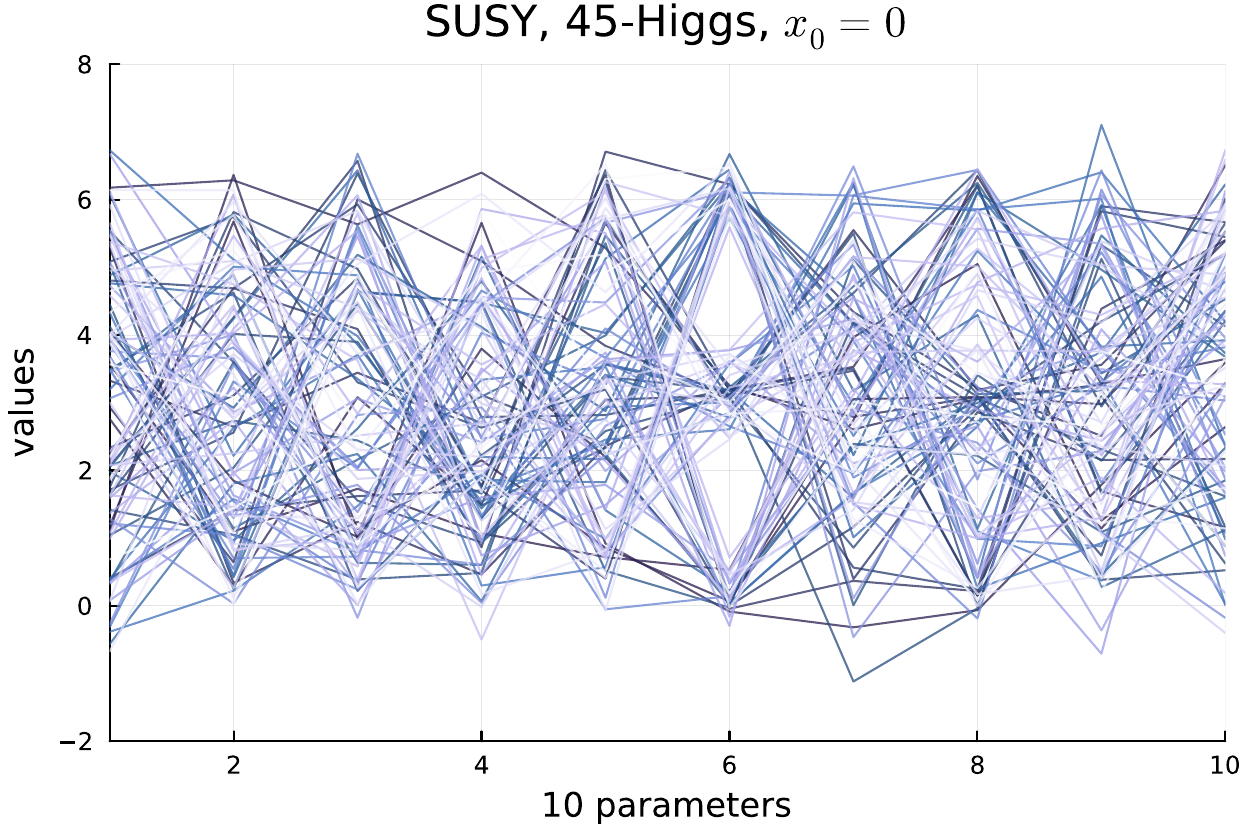}%
\includegraphics[width=60mm]{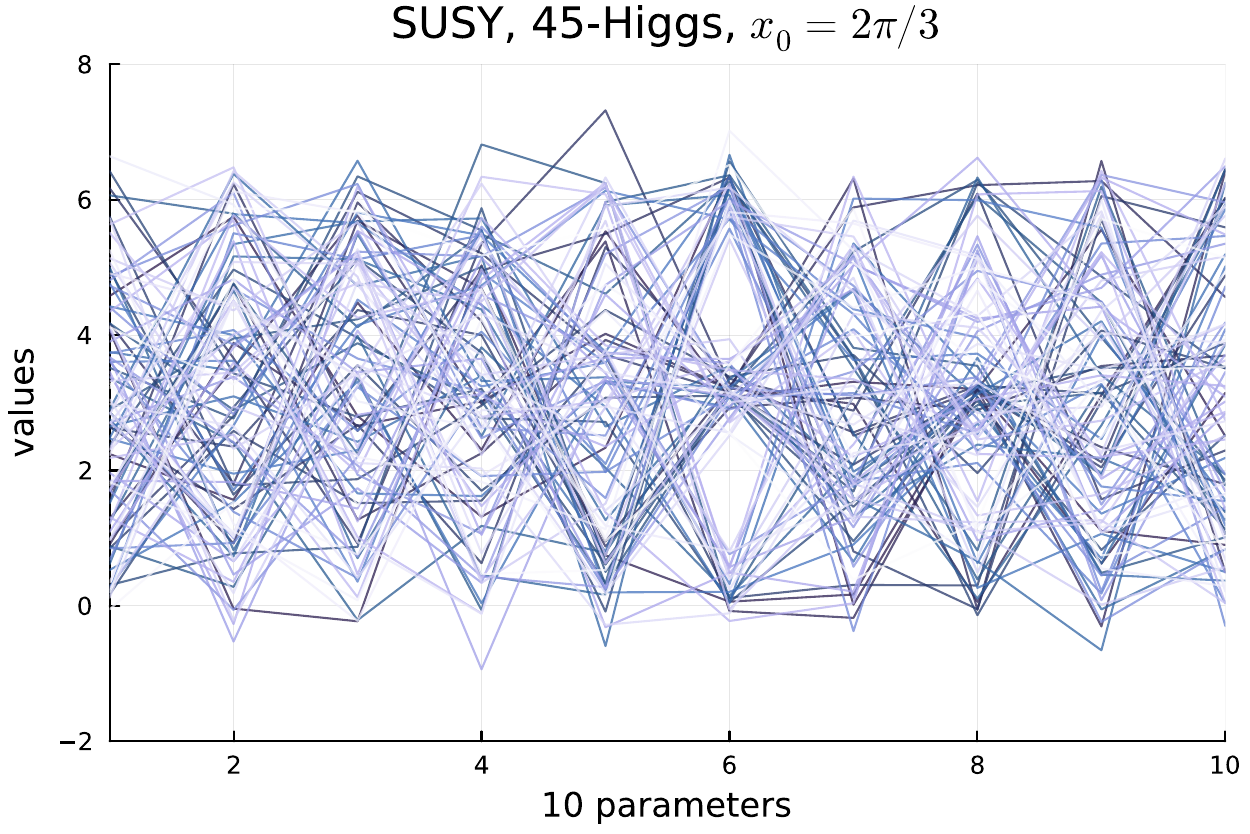}%
\includegraphics[width=60mm]{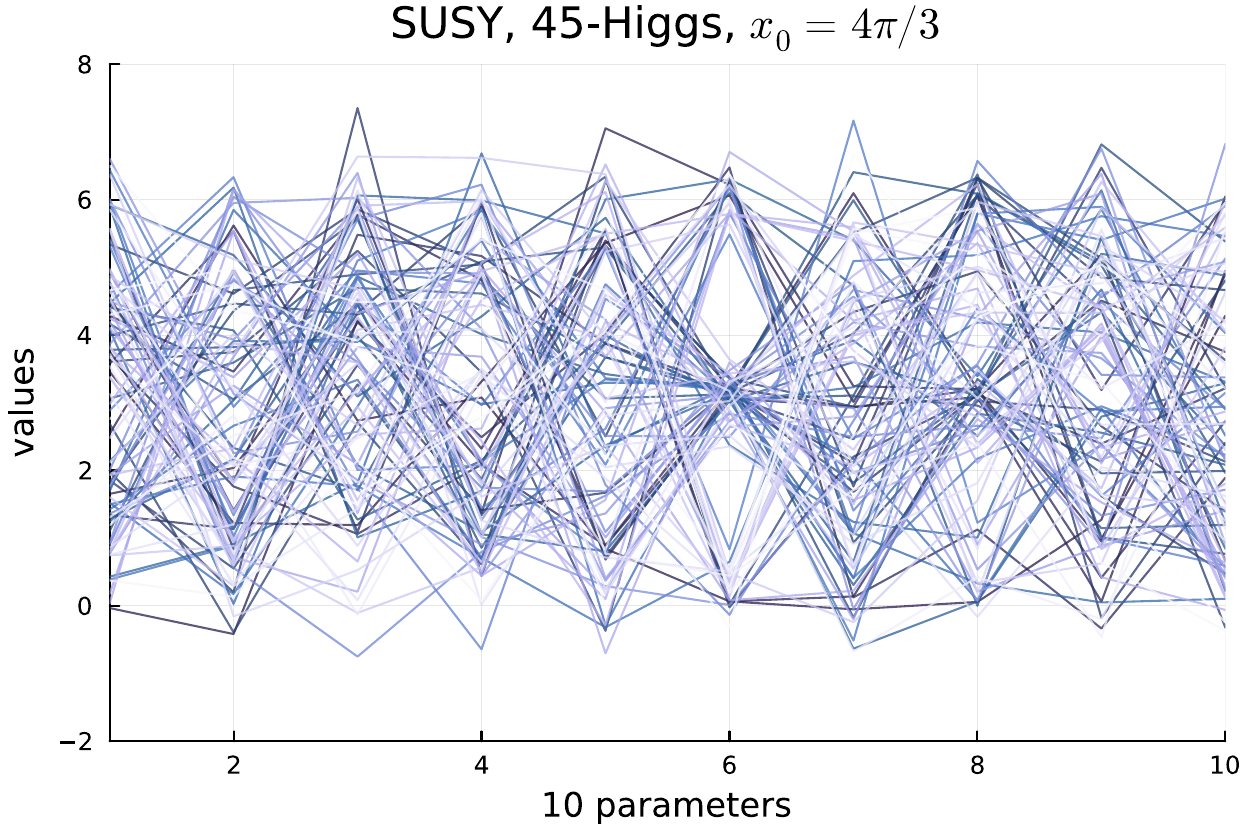}\\
\includegraphics[width=60mm]{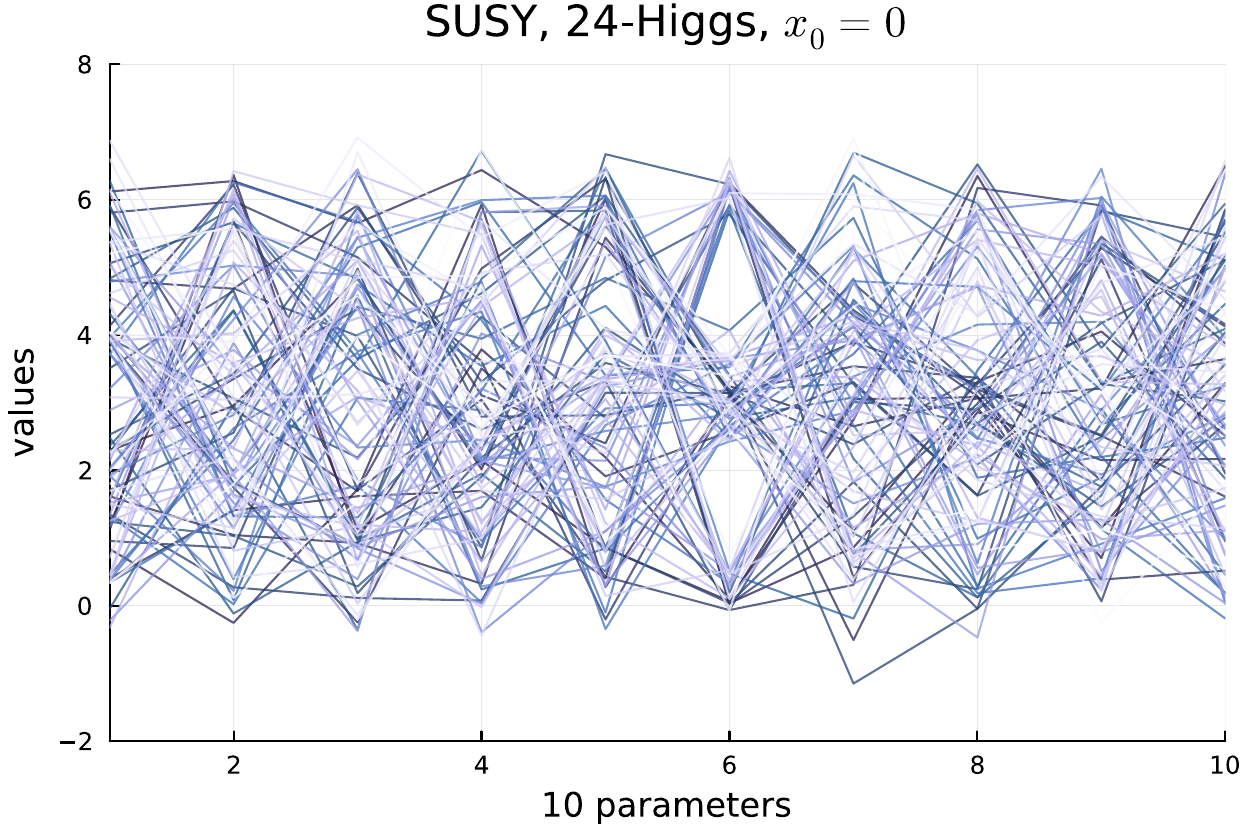}%
\includegraphics[width=60mm]{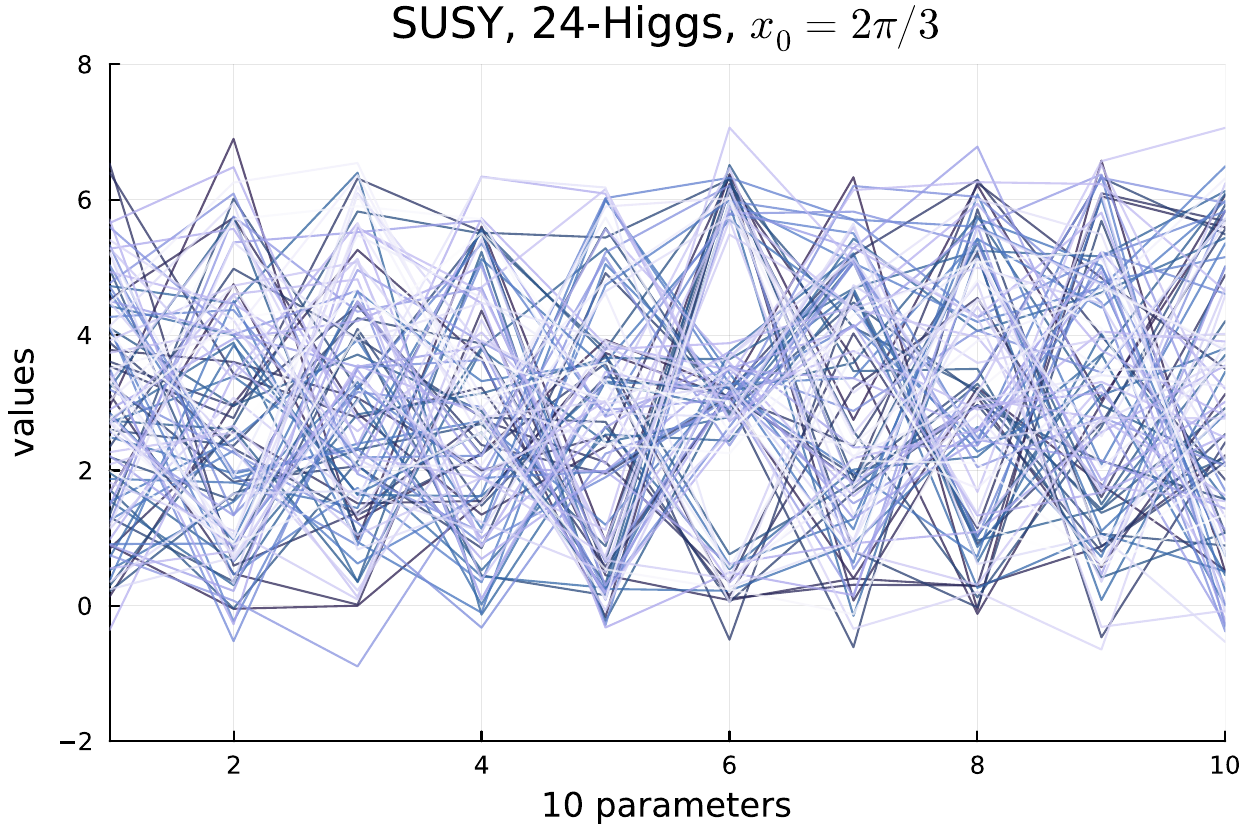}%
\includegraphics[width=60mm]{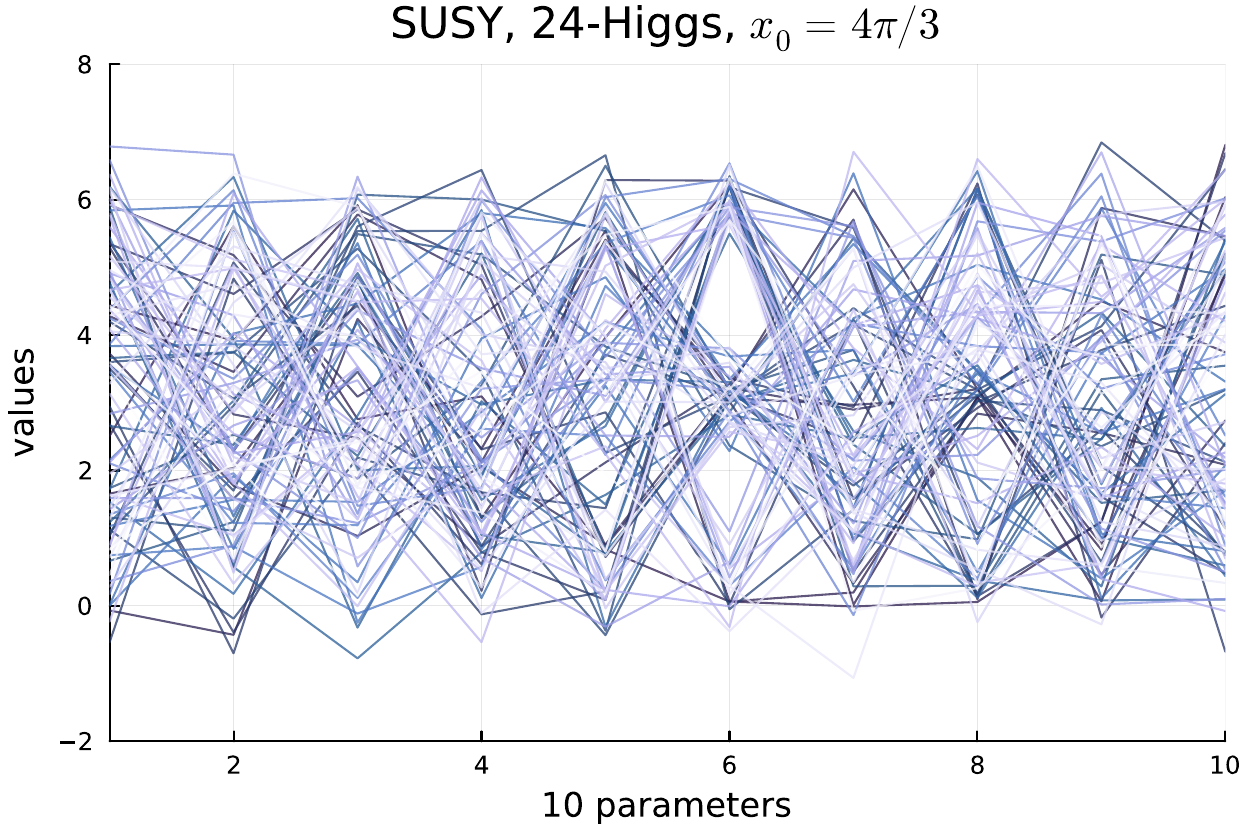}%
\caption
{\label{fig:sxVals}
Optimised configurations of ten parameters $x_1,\ldots,x_{10}$ in the supersymmetric $SU(5)$ GUT scenario.
The upper (lower) panels present the results for the 45-Higgs (24-Higgs) model. The cases where $x_0=0$, $2\pi/3$, and $4\pi/3$ are depicted on the left, middle, and right panels, respectively.
Each panel displays 100 samples representing the 100 smallest values of the loss function out of the 1024 available samples. 
Darker lines correspond to the smaller loss function values.
}
\end{figure*}

\subsection{\label{sec:SUSY_results}Good parameter values: numerical method}%

Our purpose is to investigate the parameter space of the flavour sector using the methods of machine learning, namely, sampling, optimisation and statistical analysis as described in Sec.~\ref{sec:nonSUSY_results}. 
Here we study the 45-Higgs model and 24-Higgs model of the supersymmetric $SU(5)$.
We use the same criterion of beauty specified by the loss function \eqref{eqn:Ldef}. 
The mass matrices in the supersymmetric $SU(5)$ scenario are in the same forms as the nonsupersymmetric case, \eqref{eqn:MM45} and \eqref{eqn:MM24}.
The structure of the Higgs sector is different in that there are two Higgs quintets in the supersymmetric scenario that can take different expectation values $v_d$ and $v_u=v_d\tan\beta$.
However, the fermion mass relations are concerned only with $v_d$, and the difference of $v_u$ and $v_d$ is absorbed by rescaling $y^{5u}$.
Thus in practice, the difference between the supersymmetric and nonsupersymmetric scenarios is only in the renormalisation group equations. 
Hence, we shall use the fermion mass data of the supersymmetric scenario \eqref{eqn:sGUTmass} and \eqref{eqn:sGUTCKM} at the unification scale, and repeat the numerical studies as explained in Sec.~\ref{sec:nonSUSY}.

\subsection{\label{sec:SUSY_summary}45-Higgs model vs. 24-Higgs model}%

We optimised the 10 parameters $x_1,\ldots,x_{10}$ for both 45-Higgs and 24-Higgs models, with the discrete parameter value of $x_0$ chosen to be $0$, $2\pi/3$ and $4\pi/3$. 
The loss function to be minimised\footnote{
The definition of the loss function \eqref{eqn:Ldef} differs from the one used in \cite{Kawai:2024pws} by normalisation. 
We use \eqref{eqn:Ldef} here as it is more convenient when the generalised model of Sec.~\ref{sec:ymodel} is discussed. 
So far as the 45-Higgs and 24-Higgs model are concerned this redefinition does not alter the conclusion of Ref.~\cite{Kawai:2024pws}.
}
is \eqref{eqn:Ldef}, where $M_d$, $M_e^T$ and $M_5$ are given by \eqref{eqn:MM45} for the 45-Higgs model and \eqref{eqn:MM24} for the 24-Higgs model.

In each case, we collected 1024 numerical samples (i.e., $2 \times 3 = 6$ sets).
Fig.~\ref{fig:sLhist} illustrates the distribution of the minimised values of the loss function.
We conducted optimisation with $N_{\rm iter}=10^6$ iteration steps, and the minimised loss function values were obtained by averaging over the last 100 steps.
It is observed that for all cases of $x_0 = 0$, $2\pi/3$, and $4\pi/3$, the minimised values of the loss function for the 24-Higgs model are distributed at smaller values compared to those of the 45-Higgs model. 
This suggests that the 24-Higgs model is more beautiful than the 45-Higgs model.
The results of the three cases, $x_0 = 0$, $2\pi/3$, $4\pi/3$, are observed to be similar and these three values of $x_0$ are deemed equally beautiful.

Fig.~\ref{fig:sxVals} illustrates the optimised configurations of the ten parameters $x_1,\ldots,x_{10}$ that minimise the loss function after $N_{\rm iter}=10^6$ iteration steps.
Each panel displays samples representing the top 100 most beautiful configurations out of 1024, i.e. yielding the 100 smallest loss function values after optimisation.
Darker blue indicates a smaller value of the minimised loss function, thereby representing the favoured parameter configurations according to our criteria of beauty.

\section{\label{sec:ymodel}One-parameter generalisation}

\begin{figure*}
\includegraphics[width=60mm]{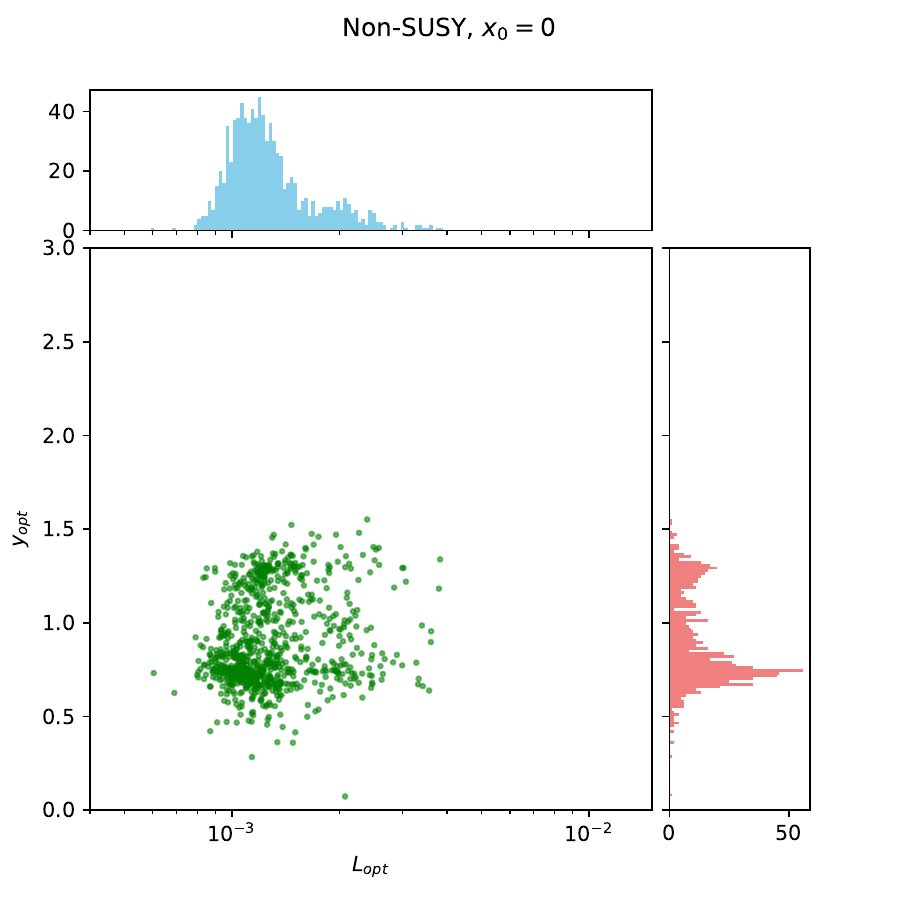}%
\includegraphics[width=60mm]{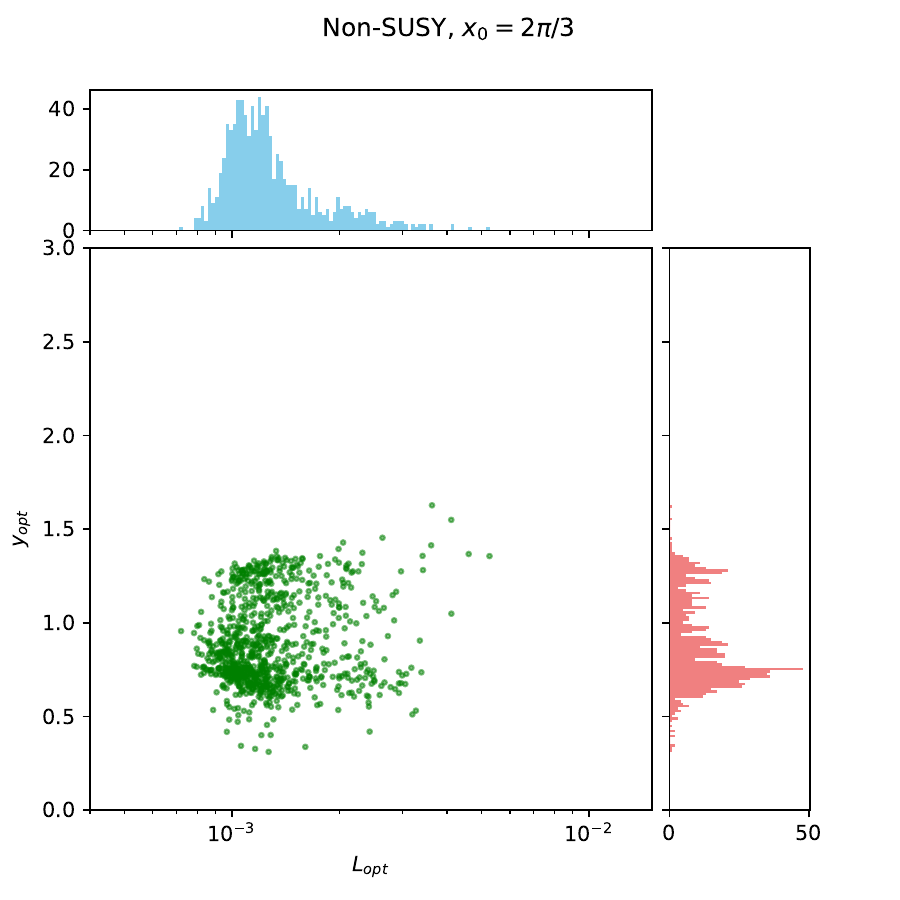}%
\includegraphics[width=60mm]{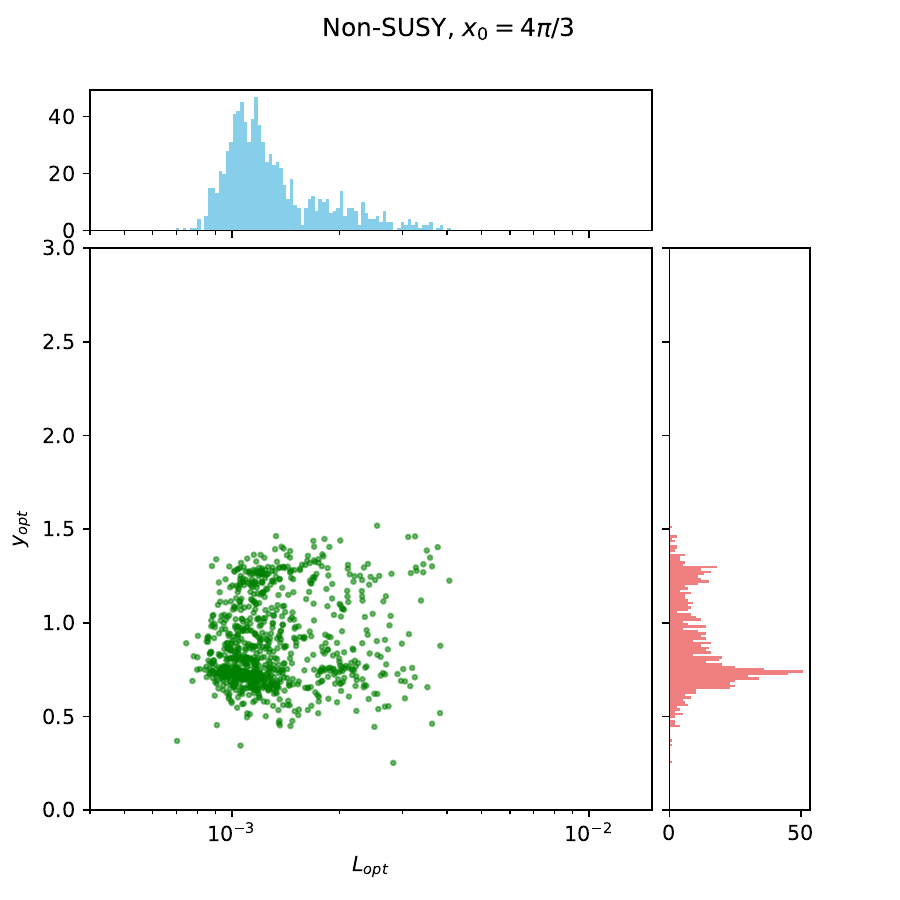}%
\caption
{\label{fig:nLy}
Optimised values of the loss function and parameter $y$, for $x_0=0$ (left panel), $x_0=2\pi/3$ (middle) and $x_0=4\pi/3$ (right), in the generalised model of nonsupersymmetric SU(5) GUT.
Optimisation was conducted for iteration steps $N_{\rm iter}=10^6$ and 1024 samples were collected for each case.
A small fraction ($<10\%$) of samples resulted in $y<0$ and were discarded.
The optimised (minimised) loss function values are the average of the last 100 steps.
The distribution of the optimised values for the loss function (light blue) and $y$ (pink) are also presented in subplots. 
}
\end{figure*}

\subsection{\label{sec:ymodel_intro}Generalisation of 45-Higgs and 24-Higgs models}%

The original $SU(5)$ GUT proposed by Georgi and Glashow \cite{Georgi:1974sy} presents an elegant framework, although it is in conflict with the observed fermion masses.
We discussed above the two notable improved models: the 45 Higgs model by Georgi and Jarlskog \cite{Georgi:1979df} and the 24 Higgs model by Ellis and Gaillard \cite{Ellis:1979fg}. 
These improved models, while successfully accommodate the observed fermion mass spectrum, involve numerous additional degrees of freedom that compromise their predictability. 
In the preceding sections, we examined these models, both for nonsupersymmetric and supersymmetric scenarios, and discovered that the 24-Higgs model exhibits a more proximity to the original Georgi-Glashow model, as determined by the criteria of proximity defined in terms of the loss function. 
In this section we pose another question, whether there exists a model that can approach closer to the Georgi-Glashow model than the 24-Higgs model. 
Machine learning techniques are found to be useful for this problem.

We saw that the mass matrices in the 45 Higgs were
\begin{align}
	M_d= M_5 + M_{45},\quad M_e^T = M_5-3 M_{45},
\end{align}
and those of the 24 Higgs model were
\begin{align}
	M_d= M_5 + 2M_{24},\quad M_e^T = M_5-3 M_{24}.
\end{align}
These are in the form of
\begin{align}\label{eqn:Mdymodel}
	M_d= M_5 + aM,\quad M_e^T = M_5 -bM,
\end{align}
if the matrices $M_{45}$ and $M_{24}$ are denoted by $M$.
Two paremeters $a$ and $b$ are assumed to be real and positive.
The 45-Higgs model is realised by $(a,b) = (1,3)$, and the 24-Higgs model is realised by $(a,b)=(2,3)$, as special cases of this generalised model.

We consider the same loss function \eqref{eqn:Ldef}. 
Using \eqref{eqn:Mdymodel}, it can be written
\begin{align}\label{eqn:Lymodel}
	L&= \left|\frac{\det(M_d-M_e^T)}{\det(M_5)}\right|\crcr
&=(a+b)^3\left|\frac{\det(M_d-M_e^T)}{\det(bM_d+aM_e^T)}\right|\crcr
&=(1+y)^3\left|\frac{\det(M_d-M_e^T)}{\det(yM_d+M_e^T)}\right|,
\end{align}
where $y=b/a$ is a real, continuous parameter.
Given that $a$ and $b$ are both positive, $y=b/a$ must also be positive.
The generalised model represents the 45-Higgs model when $y=3$ and the 24-Higgs model when $y=1.5$.
To justify this model, one may consider, for instance, a situation where the $\bm{\overline{45}}$ representation Higgs and the higher-dimensional operator coexist, and both \eqref{eqn:H45lag} and \eqref{eqn:H24lag} terms contribute to the Lagrangian.
If the Yukawa matrices of these two terms are proportional, the model reduces to our case\footnote{
Here the alignment of the Yukawa matrices of different origins is not motivated by underlying physics; it is a simplifying ansatz to keep the model tractable. 
}.

Our standard of beauty is the same as before.
We minimise the loss function \eqref{eqn:Lymodel}, by optimising eleven parameters, $x_1,\ldots,x_{10}$ and $y$.
Our aim here is to find the optimal value of $y$ by machine learning.

\begin{figure*}
\includegraphics[width=60mm]{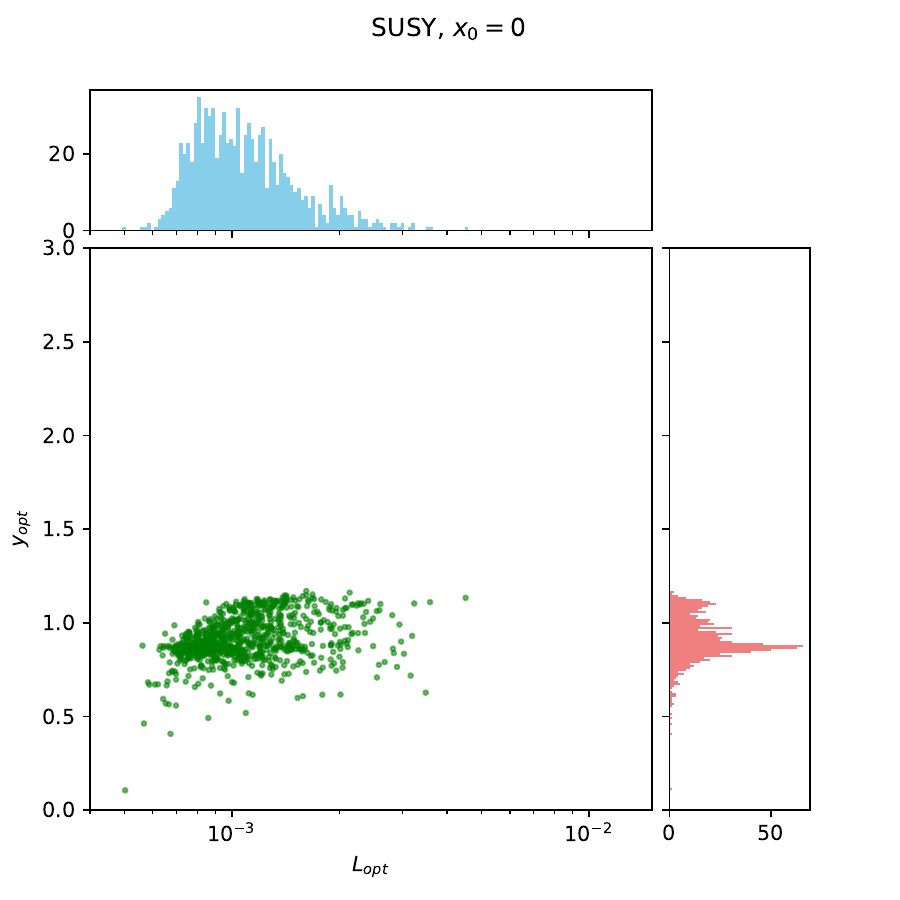}%
\includegraphics[width=60mm]{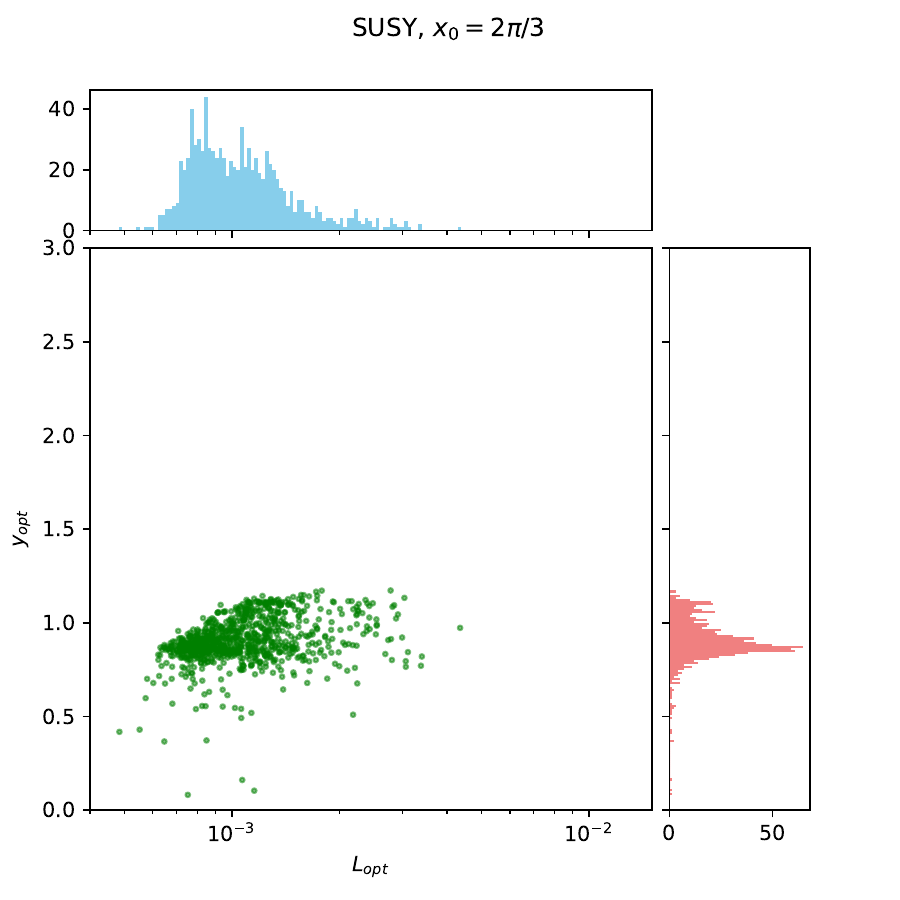}%
\includegraphics[width=60mm]{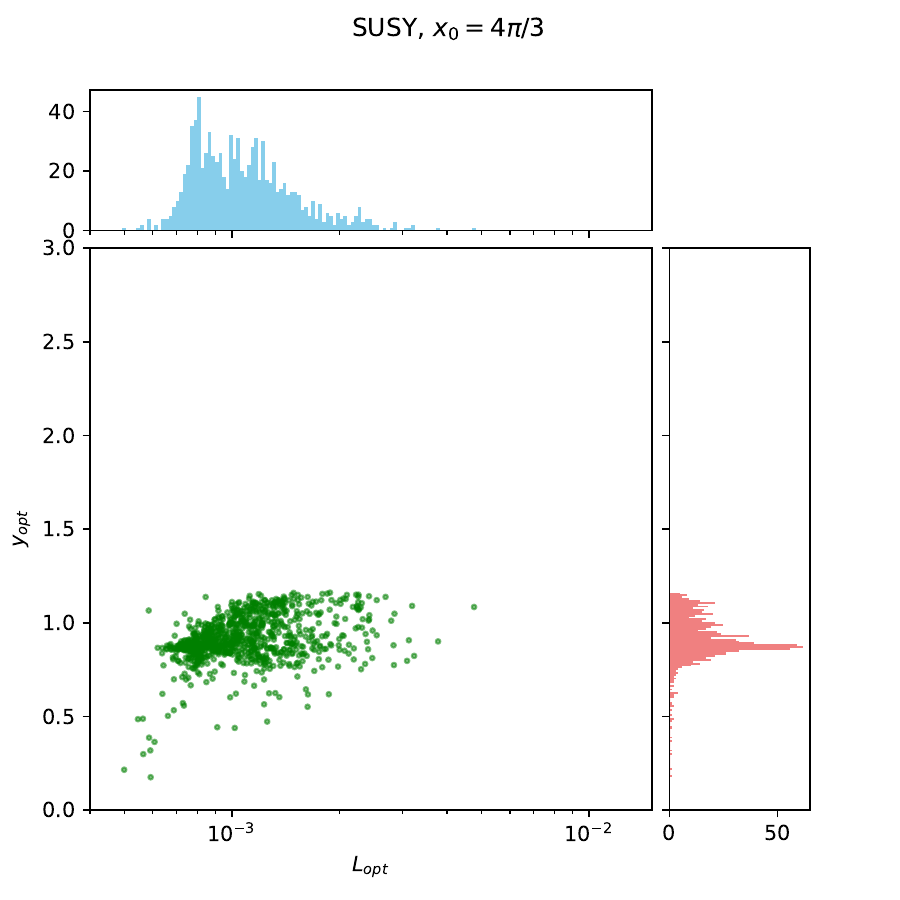}%
\caption
{\label{fig:sLy}
Optimised values of the loss function and parameter $y$, for $x_0=0$ (left panel), $x_0=2\pi/3$ (middle) and $x_0=4\pi/3$ (right), in the generalised model of supersymmetric SU(5) GUT.
Optimisation was conducted for iteration steps $N_{\rm iter}=10^6$ and 1024 samples were collected for each case.
A small fraction ($<10\%$) of samples resulted in $y<0$ and were discarded.
The optimised (minimised) loss function values are the average of the last 100 steps.
The distribution of the optimised values for the loss function (light blue) and $y$ (pink) are also presented in subplots. 
}
\end{figure*}

\subsection{\label{sec:ymodel_nonSUSY}Nonsupersymmetric scenario}%

In this subsection, we consider the nonsupersymmetric scenario.
Similar to the analysis presented in Sec.~\ref{sec:nonSUSY}, we examine the nonsupersymmetric $SU(5)$ model depicted in Table~\ref{tab:table1} and utilise the flavour data provided in Eqs.~\eqref{eqn:nGUTmass} and~\eqref{eqn:nGUTCKM} at the unification scale.
We investigate three distinct cases, each with a different value for $x_0$: $x_0=0$, $2\pi/3$, and $4\pi/3$.
To begin, we generate randomly initialised data consisting of eleven parameters, each within the range $0\leq x_1,\ldots,x_{10}<2\pi$ and $0<y\leq 3$. 
Subsequently, we minimise the loss function through a series of iterations, with a total of $N_{\rm iter}=10^6$ iterations.
This process enables us to collect 1024 samples of numerically optimised parameter sets.

Fig.~\ref{fig:nLy} illustrates the minimised loss function values (averaged over the last 100 steps of the total $N_{\rm iter}=10^6$ iterations) and the corresponding optimised parameter $y$ values.
Among the 1024 samples, $\lesssim 10\%$ deviate to $y<0$ (this numerical artefact arises from a feature of the optimisation scheme and our definition of the loss function, Eqn.~\eqref{eqn:Lymodel}, which becomes zero at $y=-1$); 
these samples have been discarded.
The distribution for $y$ exhibits a peak around $y\approx 0.75$. 
This value differs from the 45-Higgs model ($y=3$) or the 24-Higgs model ($y=1.5$), but it aligns with the results presented in Sec.~\ref{sec:nonSUSY_summary}, which indicate the preference for the 24-Higgs model over the 45-Higgs model, as $y=1.5$ is closer to $y\approx 0.75$.
The group-theoretical reason for this value, $y\approx 0.75$, as suggested by the machine learning approach, is unknown to us.
Fig.~\ref{fig:nLy} also indicates the appearance of another peak at $y\approx 1.3$.
All three cases of $x_0=0$, $2\pi/3$, and $4\pi/3$ give similar results, indicating these three values are equally beautiful by our criteria. 

\subsection{\label{sec:ymodel_SUSY}Supersymmetric scenario}%

The supersymmetric version of the generalised model is similar to above, except that the fermion mass data of the supersymmetric $SU(5)$, Eqs.~\eqref{eqn:sGUTmass} and~\eqref{eqn:sGUTCKM}, are utilised at the unification scale.
Numerical optimisation of eleven parameters, $x_1,\ldots,x_{10}$ and $y$, is carried out with randomly generated initial values from the uniform distribution in the range $0\leq x_1,\ldots,x_{10}<2\pi$ and $0<y\leq 3$. 
We collected 1024 samples for each case of $x_0=0$, $2\pi/3$, and $4\pi/3$.

Fig,~\ref{fig:sLy} illustrates the minimised loss function values (averaged over the last 100 steps of the total $N_{\rm iter}=10^6$ iterations) and the corresponding optimised parameter $y$ values.
A small fraction ($\lesssim 10\%$) of samples that resulted in negative values of $y$ have been discarded as numerical artefacts.
The distribution for $y$ exhibits a peak around $y\approx 0.85$, which differs from the values representing the 45-Higgs model ($y=3$) or the 24-Higgs model ($y=1.5$).
This value is nevertheless consistent with the results of Sec.~\ref{sec:SUSY_summary}, as it is closer to the 24-Higgs model ($y=1.5$) than the 45-Higgs model ($y=3$).
The group-theoretical reason for this optimal value, $y\approx 0.85$, is not known to us.
There also appears to be a second peak at $y\approx 1.2$. 
The three values of $x_0$, i.e. $x_0=0$, $2\pi/3$, and $4\pi/3$, give similar results and these values are deemed equally beautiful by our criteria.

\section{\label{sec:final}Final remarks}

In this paper, we investigated the flavour sector of the SU(5) GUT using machine learning techniques.
Phenomenologically viable SU(5) GUT models necessarily involve an enlarged Yukawa sector for which, due to the curse of dimensionality, a comprehensive parameter scan is impractical.
We showed that machine learning techniques (gradient-based optimisation of a loss function combined with parameter sampling) can be applied to analyse this type of problem. 
We focused on two well-known models, one involving the 45-dimensional Higgs field (45-Higgs model) and the other involving the 24-dimensional Higgs field (24-Higgs model), and compared these models in terms of a measure of favourability which we call {\em beauty}.
Here we defined the beauty as the proximity to the original Georgi-Glashow model of SU(5) GUT.
We first analysed the nonsupersymmetric model in which the grand unification is accomplished with the assistance of vectorlike fermions, then moved on to analyse the supersymmetric model.
In both nonsupersymmetric and supersymmetric scenarios, our numerical studies indicated that the 24-Higgs model is closer to the Georgi-Glashow model. This implies that the 24-Higgs model is more beautiful than the 45-Higgs model.

Subsequently, we proposed a novel model that incorporates a one-parameter family, denoted as $y$, thereby extending the scope of these two models.
This model may be considered representing the $SU(5)$ GUT when the 45-Higgs and 24-Higgs correction mechanisms are both present and their Yukawa matrices are aligned.
Through a numerical optimisation process, we determined that the optimal value of $y$ is approximately 0.75 in the nonsupersymmetric scenario, while it assumes a slightly higher value of approximately 0.85 in the supersymmetric scenario.
Furthermore, we identified a secondary peak associated with the primary peak. In the nonsupersymmetric case, this secondary peak is characterised by a value of $y$ approximately equal to 1.3, whereas in the supersymmetric case, it is associated with a value of $y$ approximately equal to 1.2.
These optimal values for $y$ emerged empirically by numerics.
We also carried out consistency checks (e.g. different choices of parameters) and confirmed those peaks. 
At the moment we do not have good interpretations on these peak values in terms of underlying physics.

The novel analysis presented here exemplifies the efficacy of machine learning techniques in the realm of high-energy physics beyond the Standard Model when conventional methodologies prove impractical.
We conclude by considering several new questions it raises.
Numerical optimisation indicates that certain parameter values are more favourable according to our criteria of beauty. 
As shown in Figs.~\ref{fig:nxVals} and~\ref{fig:sxVals}, certain regions are designated as favoured (dark) while others are unfavourable (blank).
It would be of great interest to explore their phenomenological implications \cite{Dorsner:2024seb}, particularly in the domains of neutrino physics baryogenesis and cosmological model construction \cite{Guth:1980zm,Arai:2011nq,Kawai:2015ryj}.
Our numerical study suggests that the 24-Higgs model, which assumes non-renormalisable interactions, is more beautiful than the 45-Higgs model that involves only renormalisable terms.
This trade-off between beauty and renormalisability is an interesting feature that may deserve further analysis.
In the generalised model introduced in Sec.~\ref{sec:ymodel}, the underlying physics behind the optimised parameter values $y \approx 0.75 - 0.85$ remains elusive.
The Yukawa term of the 45-Higgs model arise as a renormalisable interaction whereas the 24-Higgs model Yukawa-like term is higher-dimensional.
These two mechanisms can coexist, yet comprehending the interplay between two entirely distinct physical origins would require a framework beyond perturbative quantum field theory. 


\begin{acknowledgments}
This work was supported in part by the United States Department of Energy Grant 
Nos. DE-SC0012447, DE-SC0023713, and DE-SC0026347 (N.O.).
\end{acknowledgments}






%

\end{document}